\documentclass[12pt]{article}

\pdfoutput=1
\usepackage{slashed}
\usepackage{color, verbatim}
\usepackage{latexsym}
\usepackage{epsfig}
\usepackage{amsmath,accents}

\newcommand*{\dt}[1]{%
  \accentset{\mbox{\scriptsize\bfseries .}}{#1}}

\usepackage{amssymb}
\usepackage{graphicx}
\usepackage{bm}

\usepackage[font={small}]{caption}

\usepackage{cite}
\usepackage{hyperref}

\newcommand{\hhref}[1]{\href{http://arxiv.org/abs/#1}{arXiv:#1}}

\definecolor{myred}{rgb}{0.7, 0, 0}
\definecolor{myblue}{rgb}{0, 0, 0.7}
\definecolor{mygreen}{rgb}{0.04, 0.7, 0.5}
\hypersetup{colorlinks,citecolor=myred,linkcolor=myblue,urlcolor=myblue,linktocpage=true}

\makeatletter

\@addtoreset{equation}{section}
\makeatother

\newcommand{\be}{\begin{equation}}
\newcommand{\ee}{\end{equation}}
\newcommand{\bea}{\begin{eqnarray}}
\newcommand{\eea}{\end{eqnarray}}

\begin{document}

\thispagestyle{empty}

\begin{center}

\hfill MPP-2016-159  \\ 
\hfill UAB-FT-770

\begin{center}

\vspace{.5cm}

{\Large\bf A Natural origin for the LHCb anomalies}

\end{center}

\vspace{1.cm}

\textbf{
Eugenio Meg\'ias$^{\,a}$, Giuliano Panico$^{\,b}$, Oriol Pujol\`as$^{\,b}$, Mariano Quir\'os$^{\,b\,,c}$
}\\

\vspace{1.cm}
${}^a\!\!$ {\em {Max-Planck-Institut f\"ur Physik (Werner-Heisenberg-Institut), \\ F\"ohringer Ring 6, D-80805, Munich, Germany}}

\vspace{.1cm}
${}^b\!\!$ {\em {Institut de F\'{\i}sica d'Altes Energies (IFAE),\\ The Barcelona Institute of  Science and Technology (BIST),\\ Campus UAB, 08193 Bellaterra (Barcelona) Spain}}

\vspace{.1cm}
${}^c\!\!$ {\em {ICREA, Pg. Llu\'is Companys 23, 08010 Barcelona, Spain}}

\end{center}

\vspace{0.8cm}

\centerline{\bf Abstract}
\vspace{2 mm}
\begin{quote}\small
The anomalies recently found by the LHCb collaboration in $B$-meson decays seem to point towards the existence of new physics coupled non-universally to muons and electrons. We show that a beyond-the-Standard-Model dynamics with these features naturally arises in models with a warped extra-dimension that aim to solve the electroweak Hierarchy Problem. The attractiveness of our set-up is the fact that the dynamics responsible for generating the flavor anomalies is automatically present, being provided by the massive Kaluza--Klein excitations of the electroweak gauge bosons.
The flavor anomalies can be easily reproduced by assuming that the bottom and muon fields have a sizable amount of compositeness, while the electron is almost elementary. Interestingly enough, this framework correlates the flavor anomalies to a pattern of corrections in the electroweak observables and in flavor-changing processes. In particular the deviations in the bottom and muon couplings to the $Z$-boson and in $\Delta F = 2$ flavor-changing observables are predicted to be close to the present experimental bounds, and thus potentially testable in near-future experiments.

\end{quote}

\vfill

\newpage

\tableofcontents

\newpage

 \section{Introduction}
 \label{introduction}
 After the discovery of a Higgs-like particle at the LHC, the Electroweak (EW) Hierarchy Problem became arguably one of the most pressing theoretical issues in high-energy particle physics. Although several motivated extensions of the Standard Model (SM)
have been proposed to address this issue, no unambiguous experimental evidence is yet available to clearly discriminate among the various theoretical possibilities.

A straightforward way to hunt for a Beyond the SM (BSM) dynamics is to use direct searches
of new resonances at the LHC. This strategy is quite powerful in probing BSM scenarios that connect the solution of the Hierarchy Problem to the presence of new light states (around $1-2$~TeV), as for instance models based on Supersymmetry or on the composite Higgs idea.
Specific scenarios, however, could evade this strong connection and push the new states into the multi-TeV range as a consequence of either a peculiar structure of the BSM dynamics or of a mild amount of fine-tuning. In such situation, direct searches may have a hard time in successfully testing the BSM effects.

An additional, complementary approach to gain experimental information on BSM scenarios is to exploit indirect searches, which in several cases are sensitive to new-physics scales much higher than the TeV. Noticeable examples are the EW precision measurements performed at LEP and the flavor observables, in particular flavor-changing and/or CP violating processes. The latter observables, for instance, are able to probe new-physics effects suppressed
by energy scales as high as $10^5$~TeV. As can be easily understood, indirect searches can be extremely important to probe
scenarios with a high new-physics scale. However they retain their relevance also in scenarios with relatively light new resonances, since they can provide complementary constraints on the parameter space of the models.

Indirect searches can also provide some evidence for new phenomena and give some indication of the
possible BSM models that could explain them. An intriguing example are the anomalies in the $B$-meson decays recently found at the LHCb~\cite{Aaij:2014ora,Aaij:2015oid} and Belle~\cite{Abdesselam:2016llu} experiments.
In particular the ratio of branching fractions ${\rm BR}(B \rightarrow K \mu^+ \mu^-)/{\rm BR}(B \rightarrow K e^+ e^-)$, which differs
from the SM prediction at the $2.6\sigma$ level, could be suggestive of a violation of universality in the lepton sector. Several theoretical analyses already appeared in the literature trying to interpret the anomalies in a BSM perspective. The most obvious possibilities are some extensions of the SM including massive vector bosons ($Z'$)~\cite{Altmannshofer:2013foa,Niehoff:2015bfa,Descotes-Genon:2016hem},
or new resonances with mixed couplings to quarks and leptons (leptoquarks)~\cite{Kosnik:2012dj},
although it might also be due to underestimated hadronic uncertainties~\cite{Khodjamirian:2010vf}.
A shortcoming of many of these constructions is the fact that the BSM dynamics has no fundamental reason for being present, other than explaining the anomalies in $B$ meson physics. In the present work we follow a different approach: we do not add ad-hoc new states in order to fit the data, but instead we try to connect the LHCb anomaly to some BSM dynamics whose main motivation is addressing the EW Hierarchy Problem.

A natural way to achieve this aim is to consider extra-dimensional models a la Randall--Sundrum (RS)~\cite{Randall:1999ee}.
In these scenarios new massive vector bosons arise automatically as Kaluza--Klein (KK) modes of the SM gauge fields.
In particular, the KK towers of the $Z$ boson and of the photon can give rise to effective four-fermion interactions that
modify the $B$-meson decays. A necessary requirement to get large enough effects is the assumption that the
bottom quark and the muon have a non-negligible amount of compositeness~\footnote{The role of left-handed muon compositeness in deviations from lepton flavor universality has been recently considered in Ref.~\cite{Niehoff:2015bfa} in composite Higgs models with custodial protection.}, i.e.~that their wave-functions are sufficiently
localized towards the IR, such that their couplings with the KK vector modes are large.

An interesting byproduct of this set-up is the fact that additional corrections to precision observables are necessarily
generated as a consequence of the bottom and muon compositeness. Among the most relevant effects we can mention the
distortions of the bottom and the muon couplings to the $Z$-boson and the generation of $\Delta F = 2$ flavor-violating
contact operators. As we will discuss in Sect.~\ref{sec:Constraints}, in Natural scenarios that solve the Hierarchy
Problem, an explanation of the LHCb anomaly is correlated with deviations in the $Z$ couplings and $\Delta F = 2$
effects that are close to present experimental bounds. This correlation leads to a very predictive set-up, in which the main
parameters of the extra-dimensional model are almost completely fixed.

For definiteness, in our analysis we focus on a simple modification of the classical RS set-up, obtained by a soft-wall-like deformation of the AdS metric close to the IR
brane~\cite{Cabrer:2009we,Cabrer:2010si,Cabrer:2011fb,Cabrer:2011vu,Cabrer:2011mw,Carmona:2011ib,Cabrer:2011qb,Quiros:2013yaa,Megias:2015ory,Megias:2015qqh}.
This modified scenario allows to keep under control the corrections to the EW precision parameters even in the absence of
a custodial symmetry in the bulk. In this way the strong constraints on the mass scale of the KK vector fields can be
relaxed to the $\sim 2$~TeV range, thus avoiding the presence of a Little Hierarchy Problem between the EW scale and the KK scale.

The paper is organized as follows. In Sect.~\ref{model}, we present a brief overview of our model,
summarizing the main results of the existing literature. Afterwards, in Sect.~\ref{anomaly}, we analyze
the new-physics effects in the $B$-meson decays and we determine the values of the parameters that allow to fit the present experimental data.
The constraints coming from EW precision measurements, flavor observables and direct searches
are then presented in Sect.~\ref{sec:Constraints}. Finally in Sect.~\ref{conclusions},
we combine the fit of the LHCb anomaly and the experimental constraints, presenting an overall picture of the viability of our model together with a few concluding remarks.

\section{The model}
\label{model}

In this section we will present the model we will use in the rest of the work for our analysis of the $B$-meson decay anomalies.
As mentioned in the Introduction, the scenario we focus on is analogous to the usual RS set-up, the only difference being
a deformation of the background metric near the infrared (IR) boundary. The extra-dimension is thus close to AdS$_5$ near the ultraviolet (UV) brane,
whereas the conformal invariance is broken by a deformation of the metric only near the IR. This structure guarantees that the
RS explanation of the hierarchy between the UV and IR scales is still at work in our model, so that fields localized near the IR
brane (most noticeably the Higgs) ``feel'' an effective cut-off scale of TeV order. A detailed discussion of the model can be found
in Ref.~\cite{Megias:2015ory}, to which we refer the reader for further details.

In the following we will denote by $y$ the proper coordinate along the extra dimension and by $A(y)$ the warp factor defining the metric
\begin{equation}
ds^2 = e^{-2 A(y)} \eta_{\mu\nu} dx^\mu dx^\nu + dy^2\,,
\end{equation}
where $\eta_{\mu\nu} = (-1,1,1,1)$. The UV and IR branes are localized at the points $y = 0$ and $y = y_1$ respectively.
The form of the warp factor is determined by the dynamics of the scalar field $\phi$ which stabilizes the length of extra dimension. We assume its dynamics to be characterized by the superpotential
\be
W(\phi)=6k\left( 1+e^{a\phi} \right)^b\,,
\ee
where $a$ and $b$ are real (dimensionless) parameters controlling the gravitational background and $k$ is a parameter with mass dimension related to the curvature along the fifth dimension~\cite{Cabrer:2009we}. Since a one-to-one correspondence is present
between the value of the $\phi$ field and the position along the extra dimension, it is possible to trade the coordinate $y$ for the value of $\phi$.
From the superpotential we extract the explicit form of the warp factor
\be
A(\phi)=B(\phi)-B(\phi_0)\,,\qquad B(\phi)=\frac{1}{6ab}\left(\phi-\frac{e^{-a\phi}}{a}  \right)\,.
\ee
In $\phi$ coordinates the brane locations correspond to $\phi_0$ and $\phi_1$ for the UV and IR branes, respectively.
We fix throughout the paper $\phi_1=5$, while the position of the UV brane $\phi_0$ is used to fix the length of the extra dimension~\cite{Megias:2015ory}. Equivalently, setting the length of the extra dimension corresponds to fixing the value of $A_1=A(\phi_1)$ at the IR brane.

\begin{figure}[htb]
 \centering
 \includegraphics[width=7.8cm]{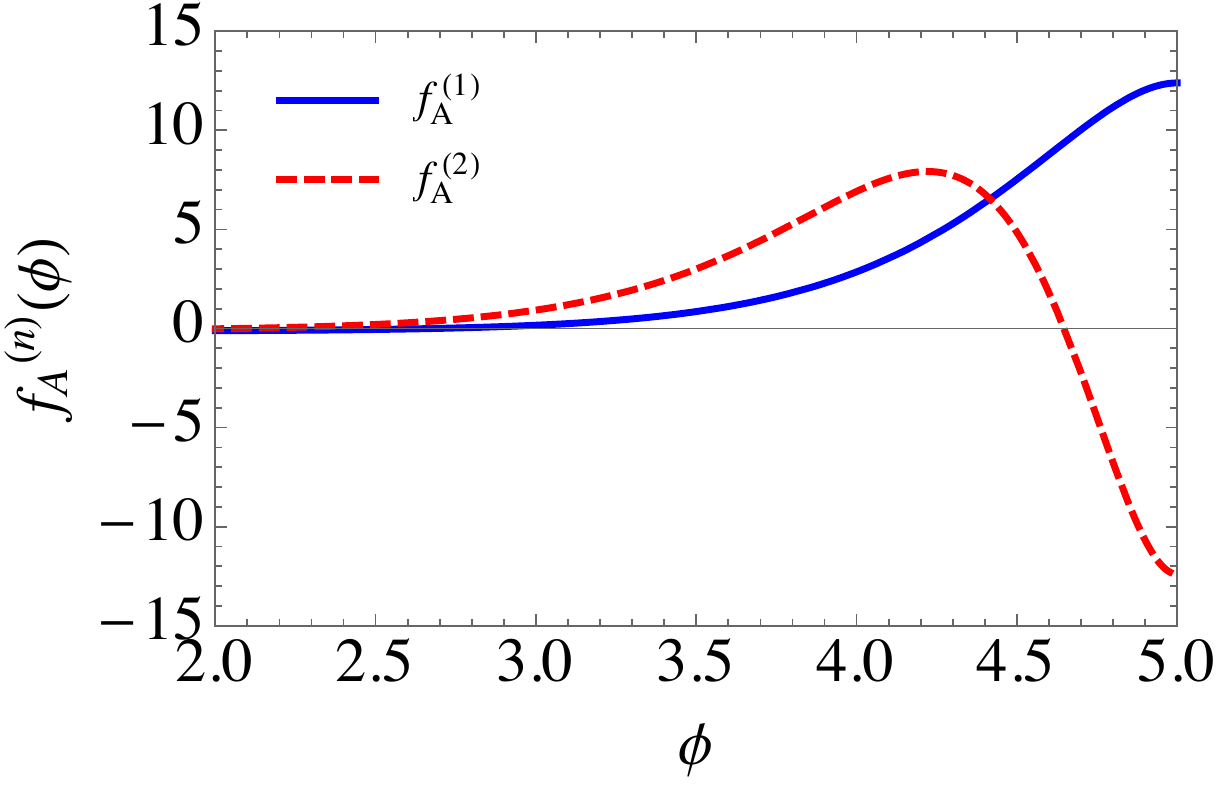}  \hfill
  \includegraphics[width=7.55cm]{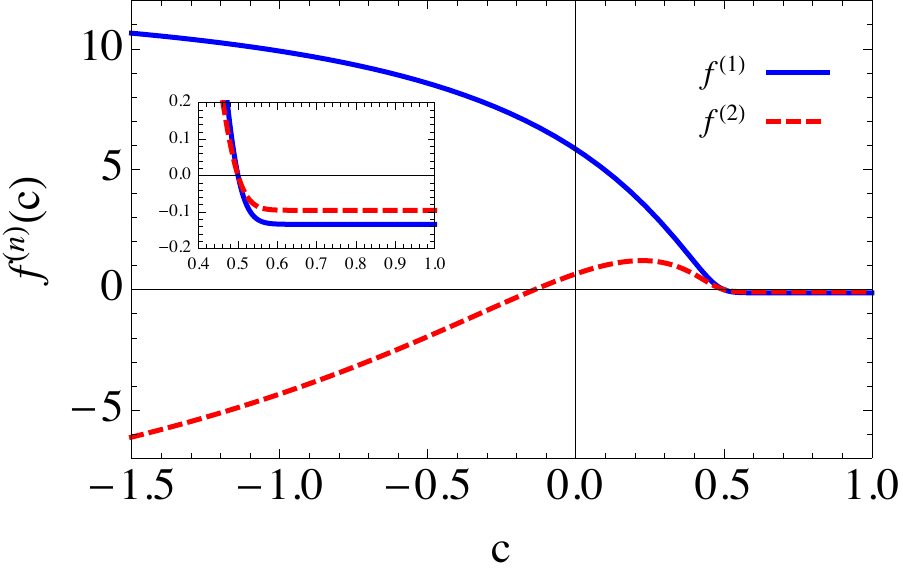} 
\caption{\it Left panel: Profiles of the gauge boson KK modes $f_A^{(n)}$ for $n=1,2$ (solid blue and dashed red lines respectively). These modes actually extend
from $\phi_0 =-8.94$ to $\phi_1=5$, but we display only the region $\phi \ge 2$ to better visualize the non-trivial behavior close to the IR brane. In the region approaching the UV brave the modes are approximately constant with values $f_A^{(1)}\simeq -0.1339$ and $f_A^{(2)}\simeq -0.0954$.
Right panel: Coupling (normalized with respect to the 4D coupling $g$) of a fermion zero-mode with the $n$-th KK gauge field, $f^{(n)}(c)$, as a function of the fermion localization
parameter $c$ (cf.~Eq.~(\ref{eq:fn})). The solid blue and dashed red lines correspond to the coupling with the first and second gauge KK modes, respectively.}
\label{KKcouplings}
\end{figure}

We assume that a five-dimensional (5D) gauge invariance is present, whose gauge group coincides with the SM one $SU(2)_L \times U(1)_Y\times SU(3)_c$. We denote the corresponding gauge fields as $W_M^a(x,y)$, $B_M(x,y)$, $G_M^A(x,y)$, where $M=(\mu,5)$, $a=1,2,3$
and $A=1,\dots,8$. The gauge fields can be decomposed in KK modes
as $A_\mu(x,y)= \sum_n f^{(n)}_A(y) A^{n}_\mu(x)/\sqrt{y_1}$.
The profiles $f_A^{(n)}(y)$, normalized as $\int_0^{y_1} \left(f_A^{(n)}(y)\right)^2 = y_1$, satisfy Neumann boundary conditions and bulk equations of motion
\begin{equation}
\left(m^{(n)}_A\right)^2f^{(n)}_A+\left( e^{-2A}\dot{f}^{(n)}_A \right)^{\dt{}}-M_A^2(y) f^{(n)}_A=0\,,
\label{gaugefields}
\end{equation}
where we use the notation $\dot{X}\equiv dX/dy$, while $m_A^{(n)}$ denotes the mass of the $n$-th KK mode and $M_A(y)$ is the mass term induced by the vacuum expectation value
of the Higgs after EW symmetry breaking (EWSB). For a Higgs propagating in the bulk, as we will assume in the following, the mass terms $M_A(y)$ has the form 
\begin{equation}
M_W(y)=\frac{g_5}{2}h(y)e^{-A(y)},\quad M_Z(y)=\frac{1}{c_W}M_W(y) \,,
\end{equation}
where $h(y)$ is the Higgs background profile and $g_5$ denotes the 5D gauge coupling corresponding to the $SU(2)_L$ subgroup. The $g_5$ coupling
is related to the usual four-dimensional (4D) one, $g$, by $g_5=g\sqrt{y_1}$.
For illustration, we plot the profiles $f_A^{(n)}$ for $n=1,2$ in the left panel of Fig.~\ref{KKcouplings} for the benchmark configuration with superpotential parameters $b=1$ and $a=0.2$ and with $A_1=35$. We will use this choice of parameters throughout the paper to derive all the numerical results. In addition we will also assume that the mass of the first gauge KK mode is of order
$m_{KK} \equiv M_1 \simeq 2$~TeV. This choice for the KK mass corresponds to the lowest value
compatible with the EW precision measurements, in particular with the bounds on the $S$ and $T$ parameters,
and with the direct LHC searches as we will discuss in Sects.~\ref{oblique} and \ref{directgluons}.

Notice that the KK modes profiles before EWSB are universal, i.e.~they do not depend on the specific gauge field they belong to. After the Higgs gets a vacuum
expectation value and the $M_A(y)$ mass turns on for the $W$ and $Z$ bosons, mild non-universal deformations of the KK wave-functions are induced. These deformations are however typically negligible for most of the computations we perform in our paper since the KK mass scale is much larger than the EW scale $v \simeq 246$~GeV.

Let us now consider the Higgs field. As we already mentioned, we assume it to be a 5D field, so that
it propagates in the bulk. Splitting the degrees of freedom into Goldstone modes $\chi(x,y)$, vacuum expectation (background) value $h(y)$ and physical fluctuations $\xi(x,y)$ we can rewrite the Higgs field as
\be
H(x,y)=\frac{1}{\sqrt{2}}e^{i\chi(x,y)}\left( \begin{array}{c}  0\\ h(y)+\xi(x,y) \end{array}   \right)\,.
\ee
EWSB is triggered by an IR brane potential, whereas additional mass terms are introduced for the Higgs in the bulk and at the UV brane. The full Higgs potential is
\be
V(H)=M^2(\phi) |H|^2+M_0|H|^2\delta(y)+\left(-M_1|H|^2 +\gamma|H|^4 \right)\delta(y-y_1)\,,
\ee
where
\be
M^2(\phi)=\alpha k \left[\alpha k-\frac{2}{3}W(\phi)  \right]\,.
\ee

\begin{figure}[t]
\centering
   \includegraphics[width=8.6cm]{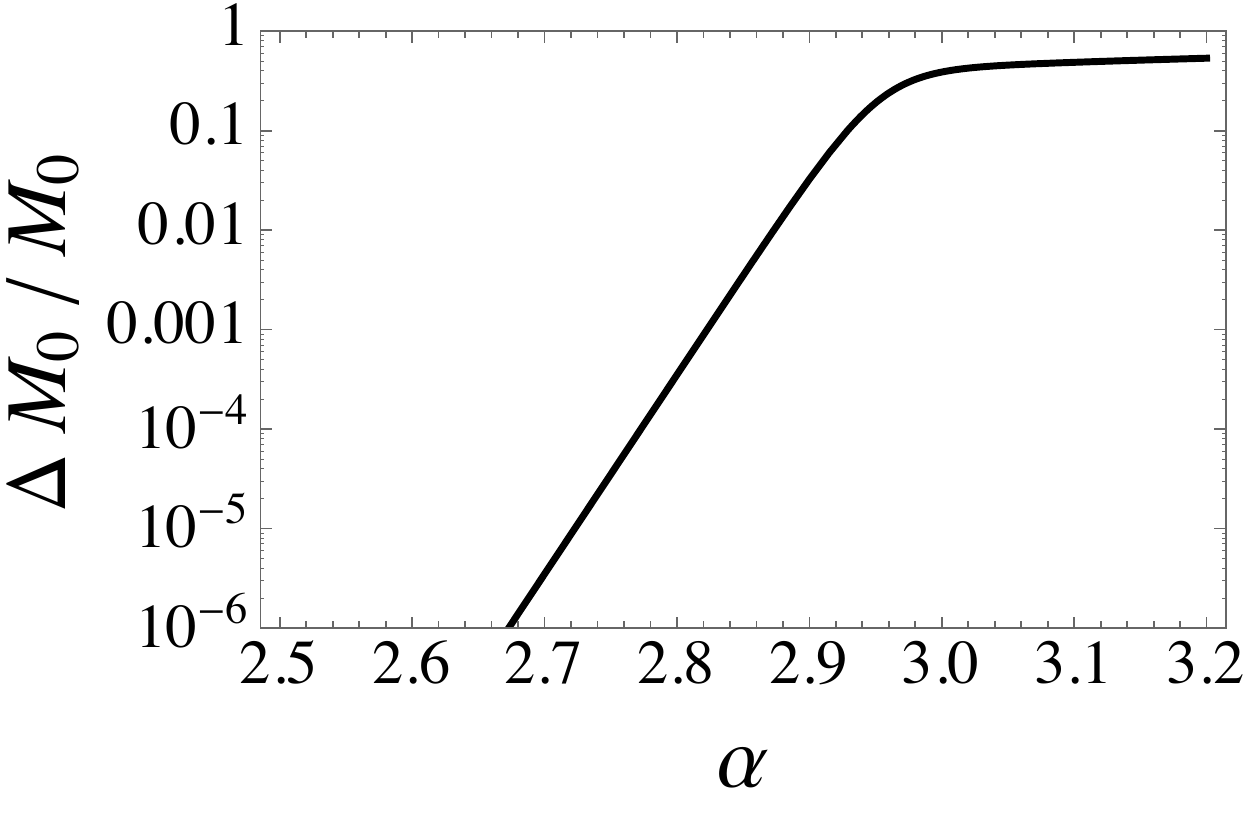} 
   \hspace{2em}
     \includegraphics[width=5.7cm]{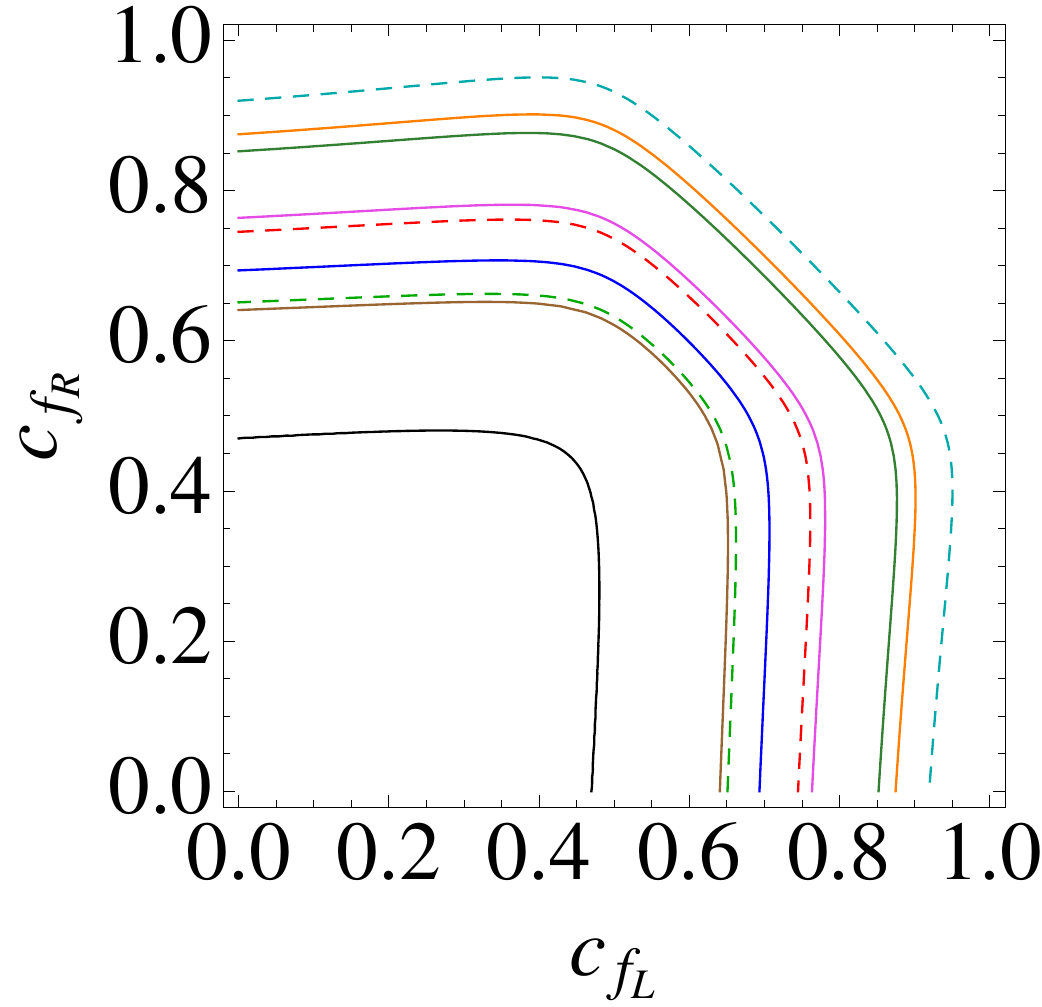}
\caption{\it Left panel: Amount of tuning $\Delta M_0/M_0$ as a function of $\alpha$. Right panel: Regions in the plane $(c_{f_L},c_{f_R})$ that accommodate the masses of the quarks and leptons. The various contours (from
the outermost to the innermost) correspond to electron, up, down, strange,  
muon, charm, tau lepton, bottom and top. As explained in the main text, in the numerical results we included the QCD running
effects for the quark masses up to the KK mass scale $m_{KK} \sim 2$ TeV.
In the parameter space region inside each contour the mass of the corresponding fermion can be obtained for values of the Yukawa couplings $\sqrt{k}\hat Y^q_{ij} \leq 3$.}
\label{fermionmasses}
\end{figure}

The dimensionless parameter $\alpha$ controls the localization of the Higgs wavefunction and can thus be
connected to the amount of tuning related to the Hierarchy Problem~\footnote{In a purely AdS metric, solving the Hierarchy Problem
requires $\alpha\geq 2$. See e.g.~Ref.~\cite{Luty:2004ye}.}.
In fact in order to ensure that the Higgs background $h(y)$ has the required exponential shape
\be
h(y) = h(0) e^{\alpha k y}
\label{higgsprofile}
\ee
a certain relation must be satisfied between the UV mass and the bulk potential, namely
$M_0=\alpha k+\Delta M_0$ with~\cite{Quiros:2013yaa}
\be
\frac{\Delta M_0}{M_0}= \frac{1}{\displaystyle \alpha k\int_0^{y_1}\!\! dy\,e^{4A-2\alpha k y}}\,.
\label{finetuning}
\ee
Obviously if $\Delta M_0$ is required to be much smaller than $M_0$ and $\alpha k$, an amount of fine-tuning of order $\Delta M_0/M_0$ is present in the model.
The integrand in the denominator of Eq.~(\ref{finetuning})
is a monotonously increasing function of $y$ and it can be easily checked that
the fine-tuning is avoided for large enough values of $\alpha$
\be
\alpha \gtrsim \alpha_1 = \frac{2 A_1}{ky_1}\,,
\ee
which correspond to localizing the Higgs background profile towards the IR brane.
For smaller values of $\alpha$ ($\alpha<\alpha_1$) some degree of fine-tuning is needed as shown
in the left panel of Fig.~\ref{fermionmasses} where we plot $\Delta M_0/M_0$ versus $\alpha$ for values of $\alpha\leq \alpha_1$ (for our choice of superpotential parameters we find $\alpha_1\simeq 3.2$). We can see that the tuning is essentially absent for $\alpha \gtrsim 2.95$, whereas it increases exponentially for smaller values of $\alpha$. 

The SM fermions are realized in our scenario as chiral zero modes of 5D fermions.
The localization of the different fermions is determined by the 5D mass terms.
The mass term for the 5D fermions
can be conveniently chosen as $M_{f_{L,R}}(y)=\mp c_{f_{L,R}} W(\phi)$
where the upper (lower) sign applies for a field with a left-handed (right-handed) zero mode~\cite{Cabrer:2011qb}.
With this convention the fermion zero modes are localized near the UV (IR) brane for $c_{f_{L,R}}>1/2$ ($c_{f_{L,R}}<1/2$).
A value $c_{f_{L,R}} < 1/2$ thus corresponds to a sizable amount of compositeness for the corresponding fermions,
whereas $c_{f_{L,R}} > 1/2$ characterizes fermions that are almost elementary.
Choosing appropriate boundary conditions we can ensure that each 5D fermion has a massless
zero mode with the appropriate chirality, whose wave-function is
\be
\psi^{(0)}_{L,R}(y,x)=\frac{e^{(2-c_{L,R})A(y)}}{\displaystyle\left(\int dy\, e^{A(1-2 c_{L,R})} \right)^{1/2}} \psi_{L,R}(x)\,.
\ee

Yukawa interactions with the Higgs are induced by the following Lagrangian
\be
\mathcal L_Y=\hat Y_{ij}^u \tilde H \bar Q_L^i u_R^j +\hat Y_{ij}^d H \bar Q_L^i d_R^j +\hat Y_{ij}^e H \bar \ell_L^i e_R^j \,,
\ee
where $\hat Y_{ij}^f$ are 5D Yukawa couplings with mass dimension $-1/2$. The dimensionless 4D Yukawa couplings are then given by
\be
Y^f_{ij}=\sqrt{k}\hat Y^f_{ij}F(c_{Q_L},c_{q_R}),\qquad F(c_L,c_R)=\frac{\displaystyle\int h e^{-(c_L+c_R)A}}{\displaystyle \left[\int h^2e^{-2A}\int e^{(1-2c_L)A} \int e^{(1-2c_R)A} \right]^{1/2}}\,.
\ee
As can be seen from the above formulae, analogously to what happens in the usual RS set-up, order-one differences in
the fermion bulk masses (i.e.~in the $c_{f_{L,R}}$ parameters) induce exponential differences in the fermion Yukawa's.
This mechanism could be interpreted as a natural way to generate the hierarchies in the fermion masses, without
any need to introduce hierarchies in the 5D Yukawa's. This set-up is usually called anarchic flavor scenario
since the 5D Yukawa's have an ``anarchic'' structure with all entries of similar order~\footnote{For the original papers
on the anarchic flavor scenario in 5D warped models see Refs.~\cite{anarchic}. For the analogous construction
in the context of the 4D interpretation of the extra-dimensional models see for instance Ref.~\cite{Panico:2015jxa}.}.
In the present work we do not fully specify the flavor structure of the model and we just consider the coefficients $c_{f_{L, R}}$
as free parameters such that for ``perturbative'' values of the Yukawa coefficients $\hat Y_{ij}^f$
the SM fermion masses and mixing angles can be reproduced. In other words, differently from the pure anarchic scenario,
in our set-up we also allow for some (mild) hierarchy in the 5D Yukawa's which should arise from some 5D flavor structure.

Since the quark and charged lepton spectrum is hierarchical, a good approximation to extract the fermion masses as a function
of the model parameters is to neglect the mixing angles (which means that $\hat Y_{ij}^f\sim 0$ for $i\neq j$).
In this way we can easily determine the range of values of $c_{f_{L,R}}$ that allow to reproduce the various fermion masses.
The numerical results, corresponding to a benchmark choice of the 5D Yukawa couplings, $\sqrt{k}\hat Y_{ii}^f=3$, are shown in the right panel of Fig.~\ref{fermionmasses}. 
In the analysis we used the values of the SM quark Yukawa couplings run at the energy scale of the
KK resonances in our model, namely $m_{KK}\sim 2$~TeV,
\begin{align}
m_t(m_{KK})&\simeq 140\ \textrm{GeV}\,,\ \
m_b(m_{KK})\simeq 2.3-2.5\ \textrm{GeV}\,, \ \
m_c(m_{KK})\simeq 0.43-0.55\ \textrm{GeV}\,, \nonumber\\
m_s(m_{KK})&\simeq 28-81\ \textrm{MeV}\,,\ \
m_u(m_{KK})\simeq 1-2\ \textrm{MeV}\,, \ \
m_d(m_{KK})\simeq 1-4\ \textrm{MeV}\,.
\end{align}

Another important ingredient we will need for our analysis is the coupling of the SM fermions with the KK modes of the gauge fields.
Before EWSB these couplings have a particularly simple form. Obviously the ones with the SM gauge bosons
coincide with the SM gauge couplings. The couplings with the massive gauge KK modes are instead universal and are
fully determined by the localization of the fermions, i.e.~by the $c_{f_{L,R}}$ parameters. 
The coupling with the $n$-th gauge KK modes, collectively denoted by $X^{n}_\mu$, can be written as
\begin{equation}
g_{f_{L,R}}^{X^n}\,X_\mu^n \bar f_{L,R}\gamma^\mu f_{L,R} \equiv g f^{(n)}(c_{f_{L,R}})X_\mu^n\, \bar f_{L,R}\gamma^\mu f_{L,R}\,,
\end{equation}
where $f_{L,R}$ are the fermion zero-mode and $g$ denotes the SM gauge coupling corresponding to the $X_\mu$ field. The coupling functions $f^{(n)}(c_{f_{L,R}})$, which encode the
overlap of the KK wave-functions of the vector bosons with the zero-mode fermion wave-functions,
are given by the following expression
\be
f^{(n)}(c)=\frac{\sqrt{y_1}}{\sqrt{\displaystyle \int_0^{y_1} \!\! \left(f_A^{(n)}(y)\right)^2}}\frac{\displaystyle \int_0^{y_1} \!\! f_A^{(n)}(y) e^{(1-2c)A(y)}}{\displaystyle \int_0^{y_1}\! e^{(1-2c)A(y)}}\,. \label{eq:fn}
\ee
The functions $f^{(n)}(c)$ for $n=1,2$ for our reference choice of the model parameters are plotted in the right panel of Fig.~\ref{KKcouplings}.
We can see that two relevant regimes are present. For a fermion zero mode localized towards the IR (namely for $c \lesssim 0.45$)
the coupling with the KK gauge fields is typically of the order of the SM gauge coupling (that is $f^{(n)}(c) \sim 1$)
and can become even stronger for $c \lesssim 0$. For fields which are almost elementary (namely for $c \gtrsim 0.5$) the coupling
instead becomes rather weak with a typical size $\sim 0.1 g$~\footnote{For $c = 1/2$ all the couplings to the KK gauge modes
exactly vanish as a consequence of the fact that the wave-functions of the fermion zero modes become flat along the extra dimension.}.
 
It is interesting to notice that, for fermions with very small compositeness, the coupling with the gauge KK's
becomes almost independent of the exact value of $c$. One can indeed check that for arbitrary $c_1,c_2\gtrsim 0.6$
the couplings differ at most at the percent level, $\left|f^{(n)}(c_1) - f^{(n)}(c_2)\right|/\left|f^{(n)}(c_1)\right| \lesssim 10^{-2}$.
This feature has very important consequences for flavor universality. Indeed gauge coupling universality is automatically ensured
for all the fermions whose localization parameters are $c_{f_{L,R}} \gtrsim 0.6$. In the following we will exploit this feature
by assuming that the first and second quark generations respect the universality condition. This implies an approximate accidental
$U(2)_{q_L} \times U(2)_{u_R} \times U(2)_{d_R}$ global flavor symmetry, which is only broken by the Yukawa couplings.
As we will see this symmetry will help in reducing dangerous flavor violating effects involving
the light quarks~\footnote{Flavor models including an approximate $U(2)$ flavor invariance for the light generations have already
been considered in the literature in the context of composite Higgs models. See for
instance Refs.~\cite{Barbieri:2012tu}.}.
As can be seen from Fig.~\ref{fermionmasses} there is no obstruction in satisfying the universality assumption
since the Yukawa couplings of the first two generations of quarks can be easily accommodated by choosing $c_{f_{L}} \sim c_{f_R} \gtrsim 0.6$.
For simplicity in our numerical analysis we will moreover choose $c_{q^{1}_L}=c_{q^{2}_L}\equiv c_{q_L}$ (where
$q^{1,2}_L$ denote the left-handed quark doublets in the first and second generation) as well as
$c_{u_R}=c_{c_R}=c_{d_R}=c_{s_R}\equiv c_{q_R}$. This choice is not strictly necessary, since any configuration
which respects $c_{f_{L,R}} \gtrsim 0.6$ would give rise to the effective $U(2)$ flavor invariance.

The results about the couplings with the KK gauge tower we presented so far are exact when the effects
of EWSB are neglected. When the Higgs gets a vacuum expectation value, however, some (possibly non-universal)
distortions of the couplings are generated. There are two main effects that modify the gauge couplings:
the mixing of the SM fermions with their KK modes and the mixing of the SM gauge fields with the
massive vector resonances. We postpone a detailed discussion of these effects to Sect.~\ref{sec:Constraints}.

\section{The $B\to K^*\mu^+\mu^-$ anomaly}
\label{anomaly}

As we mentioned in the Introduction, the recent LHCb measurements of the angular distributions in the decay
$B\to K^*\mu^+\mu^-$ and the $\sim 2.6\,\sigma$ deviation with respect to the SM prediction in the value of
$BR(B\to K\mu^+\mu^-)/BR(B\to K e^+e^-)\simeq 0.745^{+0.090}_{-0.074}\pm 0.036$~\cite{Aaij:2014ora}
suggest the possibility that universality deviations with respect to the Standard Model expectations could be present.
After EWSB the relevant four-fermion effective operators contributing to $\Delta F=1$ transitions
can be mapped into the basis~\cite{Buchalla:1995vs}
\be
\mathcal L_{eff}=\frac{G_F\alpha}{\sqrt{2}\,\pi} V^*_{ts}V_{tb}\sum_i C_i\,\mathcal O_i\,,
\label{efflagrangian}
\ee  
where the Wilson coefficients $C_i=C_i^{SM}+\Delta C_i$, are the sum of a SM contribution $C_i^{SM}$ and of a
new-physics one $\Delta C_i$. The sum in Eq.~(\ref{efflagrangian}) includes the operators~\footnote{Additional operators
involving the electron field can also contribute to $\Delta F = 1$ processes. In our set-up, however, we assume that
the first lepton generation is almost elementary, so that new contact interactions involving the electron are negligible.}
\begin{equation}
\begin{array}{l@{\qquad}l}
\mathcal O_9 =(\bar s_L\gamma_{\mu}  b_L)(\bar\mu \gamma^\mu\mu)\,, & \mathcal O_{10}=(\bar s_L\gamma_{\mu}  b_L)(\bar\mu \gamma^\mu\gamma_5\mu)\,,\\
\rule{0pt}{1.25em}\mathcal O_9^\prime =(\bar s_R\gamma_{\mu}  b_R)(\bar\mu \gamma^\mu\mu)\,, & \mathcal O_{10}^\prime=(\bar s_R\gamma_{\mu}  b_R)(\bar\mu \gamma^\mu\gamma_5\mu)\,.
\end{array}
\label{operators}
\end{equation}

\begin{figure}[htb]
\centering
 \includegraphics[width=.45\textwidth]{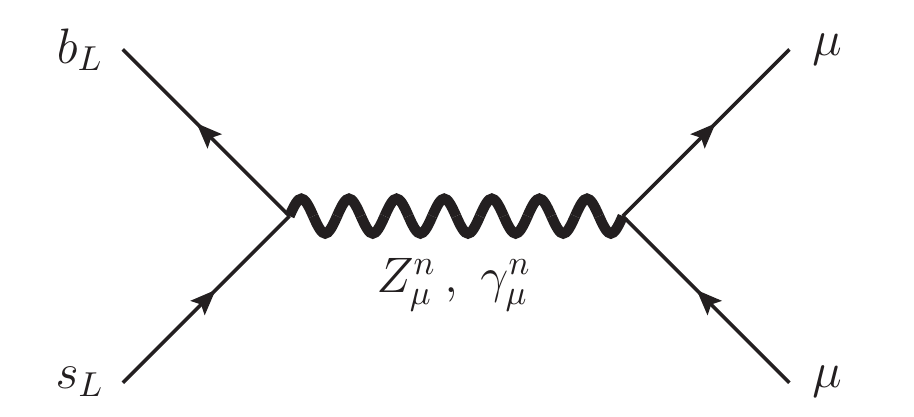}  
\caption{\it Feynman diagrams giving rise to the effective operators of Eq.~(\ref{operators})}
\label{fig:c9}
\end{figure}

In our model contact interactions involving the SM fermions can be generated through the exchange of the massive KK
modes of the gauge fields. In particular interactions that induce $\Delta F = 1$ transitions can be induced by the exchange
of the KK modes of the $Z$-boson and of the photon. The schematic structure of the diagrams giving rise to these
contributions is shown in Fig.~\ref{fig:c9}.

From the discussion in the previous section, it is easy to realize that lepton universality can be broken in our scenario,
provided that the localization of the various lepton generations is different.
In particular, if the electron is an almost elementary state ($c_{e_{L,R}} \gtrsim 0.6$) whereas the muon
has a sizable degree of compositeness, the couplings of these two fields to the gauge KK modes are
different (see the right panel of Fig.~\ref{KKcouplings}). In this case large contributions to the $\Delta F = 1$ operators in Eq.~(\ref{operators}) can be generated.

A set-up that allows to accommodate a large enough contribution to the
$\Delta F = 1$ transitions is obtained by assuming that the left-handed component
of the muon has a sizable degree of compositeness, $c_{\mu_L} \lesssim 0.5$.
Analogously we also need to assume a large compositeness for at least one of the chiralities of the bottom quark.
For the light generation quarks (as well as for the electron) we can instead assume a localization close to the UV brane,
$c_{q_{L,R}} \gtrsim 0.6$, which realizes the approximate $U(2)$ quark flavor symmetry discussed in Sect.~\ref{model}.

Before EWSB, the Lagrangian describing the interactions of the SM fermions with the vector KK modes can be written in the schematic form
\begin{align}
\mathcal L_{EW}=\sum_{X=Z,\gamma}\, \frac{X^n_\mu}{2c_W}\Big[ & g^{X^n}_{b_L}\bar b\gamma_\mu(g^{X^n}_{b_L}P_L+g^{X^n}_{b_R}P_R) b+\sum_{q=u,d,c,s}\left(g^{X^n}_{q_L}\bar q\gamma_\mu P_Lq+g^{X^n}_{q_R}\bar q\gamma_\mu P_Rq\right)\nonumber\\
&+\sum_{\ell=\mu,e}\bar\ell \gamma^\mu(g^{X^n}_{\ell_V}-g^{X^n}_{\ell_A}\gamma_5)\ell\Big]\,,
\label{lagrangianEW_noEWSB}
\end{align}
where $P_{R,L} = (1 \pm \gamma_5)/2$ denote the right and left chirality projectors. After EWSB the off-diagonal
Yukawa couplings induce a mixing between the fermions of the different generations, thus leading to flavor changing
couplings to the $Z$ and photon KK excitations. In our analysis we will assume that the unitary rotations that diagonalize
the down quark Yukawa's have the same structure of the CKM matrix, $V_{L,R}^D\sim V_{CKM}$. Due to the
larger hierarchy between the up quark-sector masses, we instead assume that the $V^U$ rotation matrices are almost equal
to the identity, $V^U_{L,R}\simeq \textbf{1}_3$~\footnote{Notice that the assumption $V^U_{L} \simeq \textbf{1}_3$ implies
that $V^D_L \simeq V_{CKM}$. However although the $V^D_R$ rotation matrix is instead not fully fixed, for definiteness we
identify it here with the CKM matrix as well. The main results we will derive depend only mildly on this assumption.}.
For the lepton sector we assume that the rotation matrices involving the charged leptons are close to the identity,
so that flavor-violating interactions with the vector KK modes are not generated. We will not specify here how the required
alignment in the lepton sector is achieved and we leave this point for future investigation.

With these assumptions, the leading flavor violating interactions with the vector KK modes have the form
\begin{align}
\mathcal L_{EW}=\sum_{X=Z,\gamma}\, \frac{X^n_\mu}{2c_W}\Big[
V_{3i}^* V_{3j}\,\bar d_i\gamma^\mu \left\{\left(g^{X^n}_{b_{L}}-\overline g^{X^n}_{L}  \right)P_L+\left(g^{X^n}_{b_R}-\overline g^{X^n}_{R}  \right)P_R\right\} d_j + {\rm h.c.} \Big]\,,\label{lagrangianEW_od}
\end{align}
where $V_{ij}$ are the CKM matrix elements, $d_i$ ($i=1,2,3$) denotes the down-type quark in the $i$-th generation
and $\overline g^{X^n}_{L,R}$ are the (universal) couplings of the first and second down quarks generation to the KK vectors.
Notice that the Lagrangian in Eq.~(\ref{lagrangianEW_od}) has a quite non-generic form, which closely resembles a minimal flavor violation structure in which all flavor-changing effects are suppressed by the CKM elements involving the third family. 
This is a direct consequence of the $U(2)$ flavor symmetry for the light
generation quarks and of the unitarity of the CKM matrix. Would the $U(2)$ symmetry be violated, potentially large flavor-changing
currents could be generated for the light quarks, and in particular $s \rightarrow d$ transitions could get sizable contributions.

The explicit expressions of the left- and right-handed couplings are
\begin{equation}
\begin{array}{l@{\qquad}l}
g^{Z^n}_{f_L} = 2(t_{3L}-Q_f s_W^2)g f^{(n)}(c_{f_L}) & g^{\gamma^n}_{f_L} = 2Q_f s_W c_W g f^{(n)}(c_{f_L})\\
\rule{0pt}{1.35em} g^{Z^n}_{f_R} = -2Q_f s_W^2g f^{(n)}(c_{f_R}) & g^{\gamma^n}_{f_R} = 2Q_f s_W c_W g f^{(n)}(c_{f_R})
\end{array}\,.
\label{couplings}
\end{equation}
From these expressions one can easily derive the vector and axial couplings, which we denote by
$g^{X^n}_{f_{V,A}} = (g^{X^n}_{f_L} \pm g^{X^n}_{f_R})/2$.

The couplings in Eqs.~(\ref{lagrangianEW_noEWSB}) and (\ref{lagrangianEW_od}) give rise to the following contributions to the
Wilson coefficients of the $\Delta F = 1$ operators:
\begin{equation}
\begin{array}{l@{\hspace{.9em}}l}
\displaystyle \Delta C_9=-\sum_{X=Z,\gamma}\,\sum_n\frac{\pi g_{\mu_V}^{X_n}\left(g^{X_n}_{b_L}-g^{X_n}_{s_L}\right)}{2\sqrt{2} G_F\alpha c_W^2 M^2_n}\,,
& \displaystyle  \Delta C_9^\prime=-\sum_{X=Z,\gamma}\,\sum_n\frac{\pi g_{\mu_V}^{X_n}\left(g^{X_n}_{b_R}-g^{X_n}_{s_R}\right)}{2\sqrt{2} G_F\alpha c_W^2 M^2_n}\,,\\
\rule{0pt}{2.em}\displaystyle \Delta C_{10}= \sum_{X=Z,\gamma}\,\sum_n\frac{\pi g_{\mu_A}^{X_n}\left(g^{X_n}_{b_L}-g^{X_n}_{s_L}\right)}{2\sqrt{2} G_F\alpha c_W^2 M^2_n}\,,
& \displaystyle \Delta C_{10}^\prime= \sum_{X=Z,\gamma}\,\sum_n\frac{\pi g_{\mu_A}^{X_n}\left(g^{X_n}_{b_R}-g^{X_n}_{s_R}\right)}{2\sqrt{2} G_F\alpha c_W^2 M^2_n} \,.
\end{array}
\label{C9}
\end{equation}
Since the size of the couplings $g_f^{X^n}$ has only a small dependence on $n$, the largest contributions to the
Wilson coefficients come from the effects of the first vector KK excitations, $Z^1_\mu$ and $\gamma^1_\mu$.
The additional contributions coming from the exchange of the higher KK modes are suppressed by their larger masses and we have checked that our results are negligibly modified by their insertion.

\begin{table}[htb]\begin{center}
\begin{tabular}{|c||c|c|c|c|}
\hline
\rule[-.4em]{0pt}{1.em}Coefficient& $\Delta C_9$& $\Delta C_9^\prime$& $\Delta C_{10}$& $\Delta C_{10}^\prime$\\
\hline
\hline
\rule[-.2em]{0pt}{1.35em}Best fit value & $-1.09$&$0.46$ &$0.56$ &$-0.25$\\
\hline
\rule[-.2em]{0pt}{1.35em}$3\sigma$ region &$[-1.67,-0.39]$ &$[-0.36,1.31]$ &$[-0.12,1.36]$ &$[-0.82,0.31]$\\
\hline
\end{tabular}
\caption{\it Fitted values for Wilson coefficients from Ref.~\cite{Descotes-Genon:2016hem}.}\label{fit}
\end{center}
\end{table}
The present experimental results allow to extract a fit of the Wilson coefficients $\Delta C_{9,10}^{(')}$.
The ranges of values compatible with the current measurements at the $3\sigma$ level are listed
in Tab.~\ref{fit}, where the fit includes, on top of the decay $B\to K^*\mu^+\mu^-$, observables from $b\to s\mu^+\mu^-$, $b\to s\gamma$ and $b\to s e^+e^-$~\cite{Descotes-Genon:2016hem}.
The experimental anomalies can be reproduced in our scenario by assuming that the left-handed component of the muon has
a sizable degree of compositeness. The right-handed component can instead be almost elementary (for definiteness
we set $c_{\mu_R}=0.7$). This set-up leaves us with three relevant free parameters which control the localization of the bottom components, $c_{b_{L,R}}$, and that of the $\mu_L$ field, $c_{\mu_L}$.

\begin{figure}[htb]
 \includegraphics[width=7.7cm]{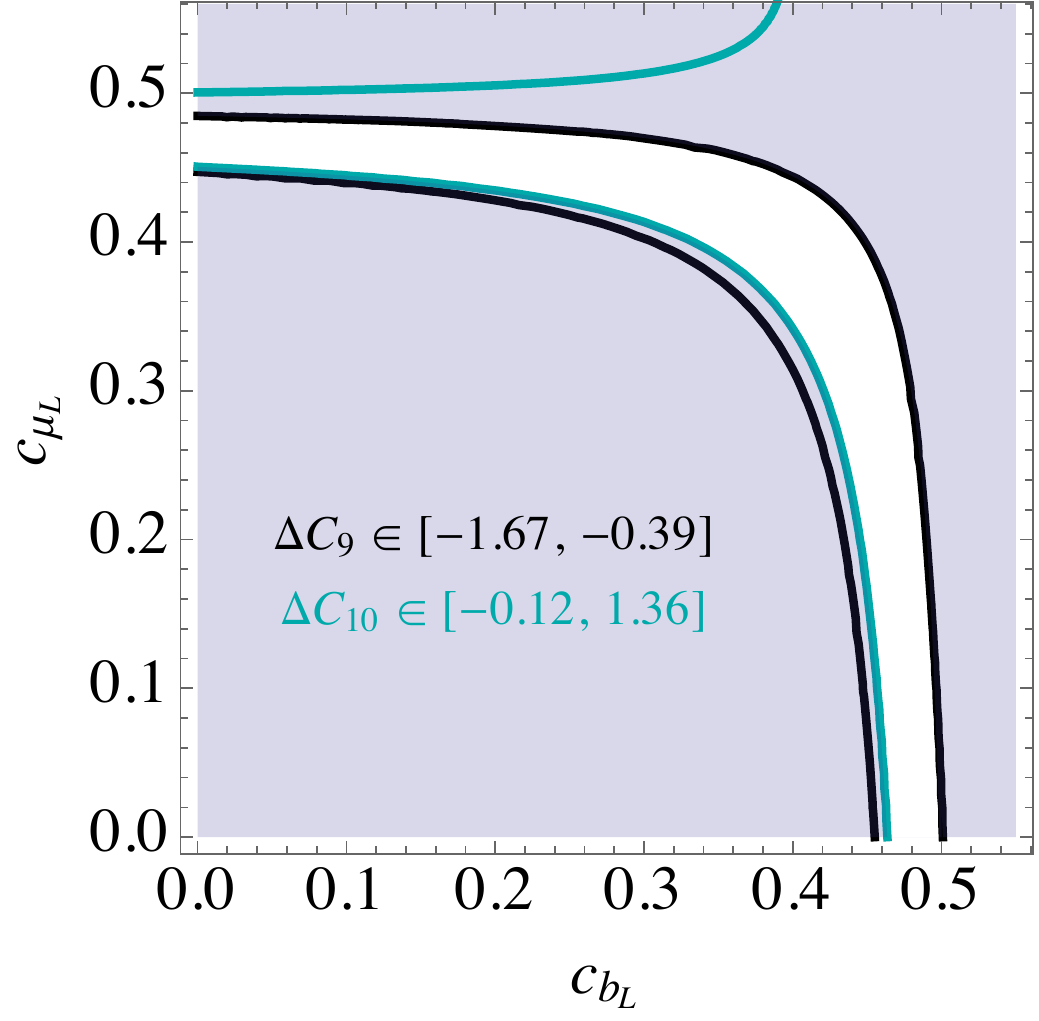}
 \hfill
 \includegraphics[width=7.7cm]{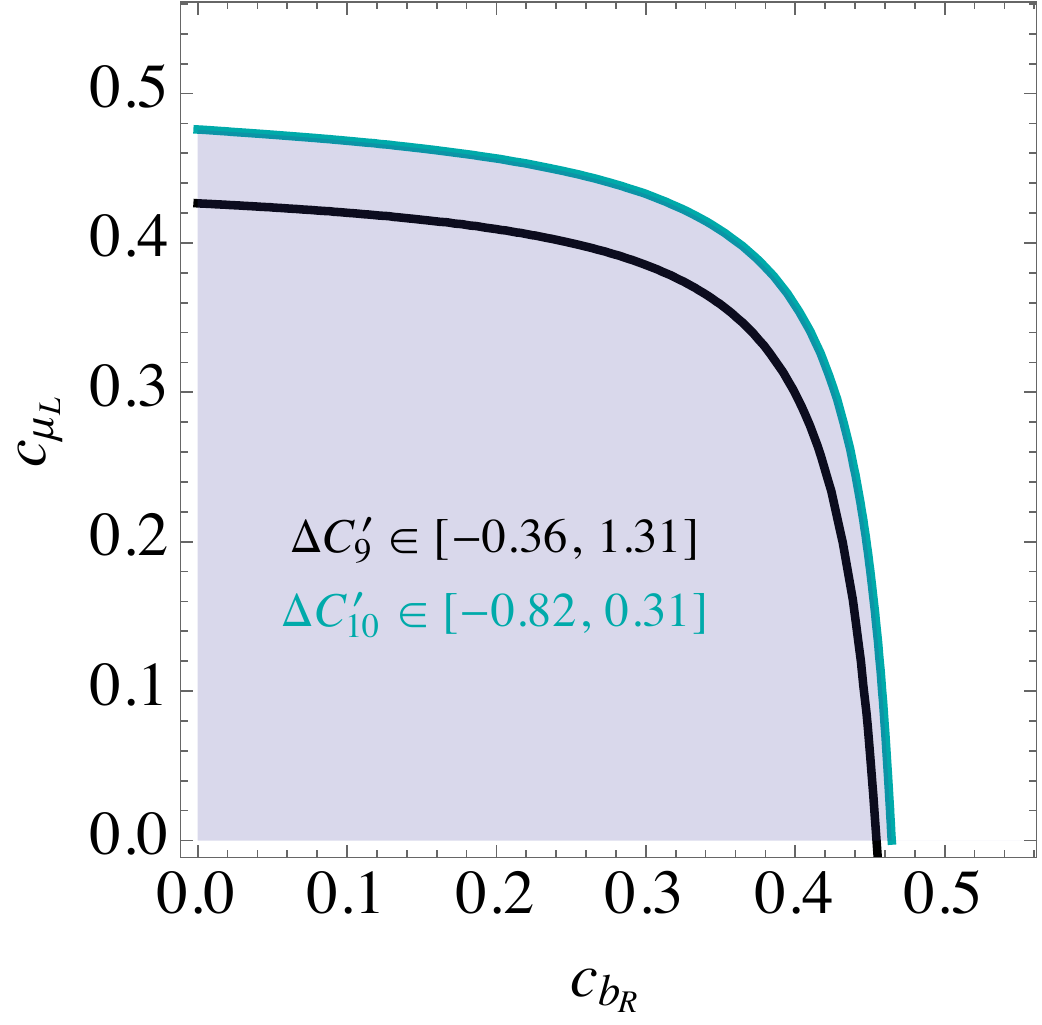}    
\caption{\it Left panel: Region in the plane $(c_{b_L},c_{\mu_L})$ compatible with the fit of $\Delta C_9$ (band between the black solid lines) and $\Delta C_{10}$ (band between the green solid lines) at the $3\sigma$ level.
Right panel: Region in the plane $(c_{b_R},c_{\mu_L})$ compatible with $\Delta C_{9}^\prime$ (the region above the black solid line) and $\Delta C_{10}^\prime$ (the region above the blue solid line). In both panels the white region denotes the points allowed
by the present data.
}
\label{LHCb}
\end{figure}

The preferred region in the parameter space is mostly determined by the value of $\Delta C_9$, which needs to
have non-vanishing new-physics contributions in order to fit the experimental results (see Tab.~\ref{fit}).
This constraint selects a relatively narrow region in the $(c_{b_L}, c_{\mu_L})$ plane. In particular, at least one of these two
parameters is required to be $\lesssim 0.45$, as can be seen from the left panel of Fig.~\ref{LHCb}.
Additional constraints on the $(c_{b_L}, c_{\mu_L})$ plane can be extracted from the bounds on $\Delta C_{10}$. These constraints, however, are quite mild and basically all points compatible with the fit of $\Delta C_9$ are also in agreement with $\Delta C_{10}$.

The $\Delta C'_{9, 10}$ Wilson coefficients, on the other hand, can be used to select a preferred region in the
$(c_{b_R}, c_{\mu_L})$ plane. As can be seen from the right panel of Fig.~\ref{LHCb}, in this case no strong constraint is found.

\section{Constraints}
\label{sec:Constraints}

As we discussed in the previous section, the exchange of gauge boson KK modes can give rise to four-fermion interactions
that modify the $B \rightarrow K^* \mu^+ \mu^-$ decay. In particular, the anomaly in the current experimental data
can be naturally explained in our model provided that the $b_L$ and $\mu_L$ fields have
a sizable degree of compositeness (i.e.~they are localized towards the IR brane). A large compositeness for the light SM fermions,
however, also implies that other precision observables can be modified. Noticeable examples are the couplings of the muon and
the bottom with the $Z$-boson and the $\Delta F = 2$ flavor observables. As we will discuss in the following, all these observables
are unavoidably linked to the generation of the ${\mathcal O}_{9,10}$ operators and the expected corrections are typically of the same order of the present experimental constraints.

\subsection{General overview}\label{sec:gen_overview}

As a preliminary step, we find it useful to perform a simple semi-quantitative analysis with the aim of giving
an overview of the various new-physics effects that can be used to put bounds on the parameter space of our model.
We postpone to the next subsections a detailed discussion and numerical analysis of each experimental constraint.

\begin{figure}[htb]
\centering
 \includegraphics[width=.4\textwidth]{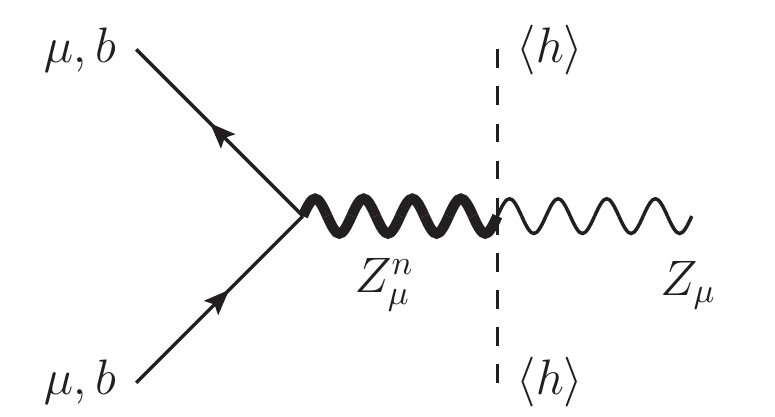}  
\caption{\it Feynman diagrams giving rise to the leading contributions to the distortion of the $Z$ couplings to the $b$-quark and to the muon.}
\label{fig:Z_coupling}
\end{figure}

A first consequence of the large bottom and muon compositeness is the fact that
sizable distortions of their couplings with the SM gauge bosons, and in particular with the $Z$, can be generated.
As we mentioned in Sect.~\ref{model}, after EWSB two effects can induce non-universal distortions of the gauge couplings.
The first effect is the mixing of the fermion zero modes with their massive KK towers. The size of these corrections
crucially depends on the localization of both the left-handed and the right-handed components of the SM fermions.
The second effect comes instead from the mixing between the $Z$-boson and its vector excitations.
In this case the corrections come from diagrams like the one shown in Fig.~\ref{fig:Z_coupling} and the distortion of
the coupling for one fermion chirality is independent of the compositeness of the other chirality.

Since the effects comings from the $Z$-boson mixing and from the fermion KK's are
controlled by different parameters, and are to a large extent independent,
they typically do not cancel each other. We can thus use the size of one of them to obtain a lower bound on
the overall size of the distortions. In this approximate analysis we will thus only consider the corrections due to the $Z$ mixing, which have a simpler structure and allow to constrain the compositeness of each fermion chirality independently.
In the detailed analysis presented in the next subsections we will also take into account the contributions coming from the fermion KK modes.

By using the notation of Eq.~(\ref{eq:fn}), we can estimate the correction to the SM fermion couplings due to the vector resonances as
\begin{equation}\label{eq:delta_Zff}
\frac{\delta g_{Zff}}{g^{SM}_{Zff}} \simeq \sum_ n f^{(n)}(c_f)\, \kappa_n \left(\frac{m_Z}{M_n}\right)^2\,,
\end{equation}
where $\kappa_n$ parametrizes the overlap of the $Z$ boson and its $n$-th KK wave-functions with the Higgs boson.
Notice that the $\kappa_n$ factors are fully determined by the geometry of the extra dimension and by the localization of the Higgs
wave-function (or equivalently by the parameter $\alpha$ in the Higgs profile, Eq.~(\ref{higgsprofile})). Since the $\kappa_n$ parameters have a small dependence on $n$,
the largest correction to the couplings comes from the mixing of the $Z$ boson with its first KK mode.
We can thus approximate the expression in Eq.~(\ref{eq:delta_Zff}) by
\begin{equation}\label{eq:delta_Zff_app}
\frac{\delta g_{Zff}}{g^{SM}_{Zff}} \simeq f^{(1)}(c_f)\, \kappa_1 \left(\frac{m_Z}{M_1}\right)^2\,.
\end{equation}
The overlap function $f^{(1)}(c_f)$ is shown in the right panel of Fig.~\ref{KKcouplings}. As can be seen from the plot, for
values $c_f \lesssim 0.4$ the overlap function can be of order $f^{(1)} \sim \mathcal O(\textrm{few})$, whereas it is suppressed
in the region $c_f \gtrsim 0.5$. The second important object in Eq.~(\ref{eq:delta_Zff_app}) is the $\kappa_1$
parameter. For typical values of $\alpha$ which correspond to a natural solution of the hierarchy problem ($\alpha \gtrsim 3$ for the chosen values of the superpotential parameters $a$ and $b$),
one finds $\kappa_1 \gtrsim 1$. Smaller values of $\kappa_1$ can be obtained by lowering the value of $\alpha$ (i.e.~by 
pushing the Higgs field towards the UV brane)~\footnote{In the limit in which the Higgs wave-function becomes flat
along the extra dimension, the Higgs VEV does not induce off-diagonal mass terms due to the orthogonality of the KK wave-functions.}.
In this case, however, an increased amount of tuning is present in the model (see Fig.~\ref{fermionmasses}).

The current experimental data put strong bounds on the deviations in the $b_L$ and $\mu_L$ couplings to the $Z$-boson,
namely $\delta g_{Zff}/g^{SM}_{Zff} \lesssim \textrm{few} \times 10^{-3}$. Substituting these constraints into Eq.~(\ref{eq:delta_Zff_app}), we find
\begin{equation}\label{eq:zcoupl_constr}
|f^{(1)}(c_{f_L})| \simeq \left|\frac{\delta g_{Z f_L f_L}}{{g^{SM}_{Z f_L f_L}}} \frac{1}{\kappa_1}\right| \left(\frac{M_1}{m_Z}\right)^2
\lesssim 1\,,
\qquad
f=\mu,b\,.
\end{equation}
As can be seen from the right panel in Fig.~\ref{KKcouplings}, the constraints in Eq.~(\ref{eq:zcoupl_constr}) roughly correspond to
$c_{\mu_L} \gtrsim 0.4$ and $c_{b_L} \gtrsim 0.4$. A comparison with the results of the fit in Fig.~\ref{LHCb} shows
that the constraints on $c_{\mu_L}$ and $c_{b_L}$ must be both saturated in order to allow a large enough contribution to ${\cal O}_9$.

We can be more quantitative by noticing that the main contribution to $\Delta C_9$ is mediated by the exchange of the
first KK modes of the $Z$-boson and of the photon (see Fig.~\ref{fig:c9}). By using the results in Eqs.~(\ref{lagrangianEW_noEWSB}) and (\ref{couplings}) we find
\begin{eqnarray}
\Delta C_9 & \simeq & - \frac{\sqrt{2} \pi}{G_F \alpha} f^{(1)}(c_{\mu_L}) f^{(1)}(c_{b_L}) \frac{1}{M_1^2}
\left(g^{SM}_{Zb_L b_L}g^{SM}_{Z\mu_L\mu_L} + g^{SM}_{\gamma b_L b_L}g^{SM}_{\gamma \mu_L\mu_L}\right)\\
& = & - \frac{\pi^2 v^2}{s_W^2} \left(1 + \frac{1}{3} \frac{s_W^2}{c_W^2}\right) f^{(1)}(c_{\mu_L}) f^{(1)}(c_{b_L}) \frac{1}{M_1^2}\,.
\end{eqnarray}
Putting this result together with Eq.~(\ref{eq:delta_Zff_app}), we finally get
\begin{equation}
\Delta C_9 \simeq - \frac{\pi^2 v^2}{s_W^2} \left(1 + \frac{1}{3} \frac{s_W^2}{c_W^2}\right)
 \left(\frac{\delta g_{Z\mu_L\mu_L}}{{g^{SM}_{Z\mu_L\mu_L}}}\right)  \left(\frac{\delta g_{Zb_L b_L}}{{g^{SM}_{Zb_L b_L}}}\right)
 \frac{1}{\kappa_1^2} \frac{M_1^2}{m_Z^4}\,.
\end{equation}
The upper value for the contributions to ${\cal O}_9$ can thus be estimated as
\begin{equation}
|\Delta C_9 | \lesssim 1\,,
\end{equation}
which is of the size required to explain the $B \rightarrow K^* \mu^+ \mu^-$ anomaly (see Tab.~\ref{fit}).

A second set of constraints can be derived from $\Delta F = 2$ flavor-changing processes. In the presence of a sizable
compositeness for the $b_L$ field, four-fermion contact operators of the form $(\bar b_L \gamma^\mu b_L) (\bar b_L \gamma_\mu b_L)$
are induced by the exchange of heavy vectors. In particular the largest contributions come from the interactions with the
KK modes of the gluons. The KK modes of the $Z$-boson and of the photon give rise to additional corrections, which are
however subleading due to the smaller gauge couplings. As we saw in Sect.~\ref{anomaly}, after EWSB
a mixing among the three quark generations is induced, whose size is parametrized by the CKM matrix elements.
Consequently the four-fermion operators involving the $b$ quark give rise to flavor changing contact interactions involving the light-quarks.

Additional four-fermion operators directly involving the light-generation quarks are also typically generated through the exchange
of KK vector bosons. These effects are however smaller than the ones coming from the third generation because in our set-up
the light quarks have a low amount of compositeness. Moreover, the $U(2)$ flavor structure we assumed for the first
two quark generations leads to a further suppression, analogous to the effect discussed in Sect.~\ref{anomaly}.

The leading contributions to the $\Delta F = 2$ operators involving four left-handed quarks have the form
\begin{equation}
{\cal O}^{LL}_{\Delta F = 2} \simeq \frac{1}{6} g_s^2 \left(f^{(1)}(c_{b_L})\right)^2 \frac{1}{M_1^2} (V^\star_{3i} V_{3j})^2
\left(\bar d_{iL} \gamma^\mu d_{jL}\right) \left(\bar d_{iL} \gamma_\mu d_{jL}\right)\,,
\end{equation}
where $g_s$ is the QCD gauge coupling and we denoted by $d_{iL}$ ($i = 1,2,3$) the down-type left-handed quark in the $i$-th generation. As can be seen from
the above expression, the flavor changing operators follow an approximate MFV structure
\begin{equation}
{\cal O}^{LL}_{\Delta F = 2} \simeq {\cal C}^{LL}\, (V^\star_{3i} V_{3j})^2
\left(\bar d_{iL} \gamma^\mu d_{jL}\right) \left(\bar d_{iL} \gamma_\mu d_{jL}\right)\,.
\end{equation}
The current bounds on the coefficient of the chirality conserving $\Delta F = 2$ operators is of the order
${\cal C}^{LL} \lesssim 1/(5\ \rm{TeV})^2$~\cite{UTfit,Isidori:2013ez} and can be translated into an upper bound on the compositeness of the $b_L$ field, namely
\begin{equation}
|f^{(1)}(c_{b_L})| \lesssim 0.8\,.
\end{equation}
This bound is of the same order of the one we derived from the constraints
on the $Z$-boson couplings in Eq.~(\ref{eq:zcoupl_constr}). This implies that, in order to explain the 
$B \rightarrow K^* \mu^+ \mu^-$ anomaly, also the corrections to the $\Delta F = 2$ processes are required to be of the order of the
present experimental bounds. Similar effects generate also four-fermion $\Delta F =2$ operators involving right-handed quarks, providing the bound $|f^{(1)}(c_{b_R})|\lesssim 0.8$.

The exchange of vector resonances gives also rise to $\Delta F = 2$ operators involving simultaneously the left- and right-handed quarks.
Among these additional operators the most strongly constrained are the ones of the
form $(\bar d_{iR} d_{jL}) (\bar d_{iL} d_{jR})$. Analogously to the left-handed operators, we find
\begin{equation}
{\cal O}^{LR}_{\Delta F = 2} \simeq g_s^2 f^{(1)}(c_{b_L}) f^{(1)}(c_{b_R}) \frac{1}{M_1^2} (V^\star_{3i} V_{3j})^2
(\bar d_{iR} d_{jL}) (\bar d_{iL} d_{jR})\,.
\end{equation}
In this case, the experimental bounds translate into a combined bound
on the amount of compositeness of the left- and right-handed bottom components. As we will discuss in Sect.~\ref{flavor},
the bounds coming from the LR operators can be more stringent than the LL and RR ones if $c_{b_L} \simeq c_{b_R}$ (see Fig.~\ref{fig:flavor}).

Finally, another set of constraints comes from the direct LHC searches for EW vector resonances decaying into a pair of muons
and for massive KK gluons decaying into top quarks.
In the presence of sizable couplings with the light quarks, the production cross section of the vector boson KK's at the LHC can
be significantly high. The current bounds allow to put some constraints on the amount of compositeness of the light quarks and of the bottom.

\subsection{Electroweak precision data}
\label{EWPD}

In this section we will discuss the main electroweak precision observables which can be used to constrain our model.
The most relevant bounds come from the universal oblique observables (encoded by the $S$, $T$
and $U$ parameters) and from the non-universal correction to the $Z$ coupling with the bottom and the muon.

\subsubsection{Oblique corrections}
\label{oblique}
To compare the model predictions with electroweak precision tests (EWPT) we will use the original $(S,T,U)$
variables defined in Ref.~\cite{Peskin:1991sw} (see also Ref.~\cite{Barbieri:2004qk}).
In our model they are given by the following expressions~\cite{Cabrer:2011fb}
\begin{align}
\alpha T & =s^2_W \frac{m_Z^2}{\rho^2}k^2 y_1\int_{0}^{y_1}
\left[1-\Omega_h(y)\right]^2e^{2A(y)-2A_1} dy\,,
\nonumber\\
\alpha S & =8c^2_Ws^2_W \frac{m_Z^2}{\rho^2}k^2 y_1\int_{0}^{y_1}
\left(1-\frac{y}{y_1}\right)\left[1-\Omega_h(y)\right]e^{2A(y)-2A_1} dy\,,\nonumber\\
\alpha U & \simeq 0\,,
\end{align}
where $\rho=ke^{-A(y_1)}$ and
\be
\Omega_h(y)=\frac{\omega(y)}{\omega(y_1)},\quad \omega(y)=\int_{0}^{y}
h^2(\bar y)e^{-2A(\bar y)} d\bar y \,.
\ee
These expressions include the leading contributions, which are due to the tree-level mixing of the
SM gauge bosons with the massive vector KK modes.
\begin{figure}[htb]
\begin{center}
 \includegraphics[width=9cm]{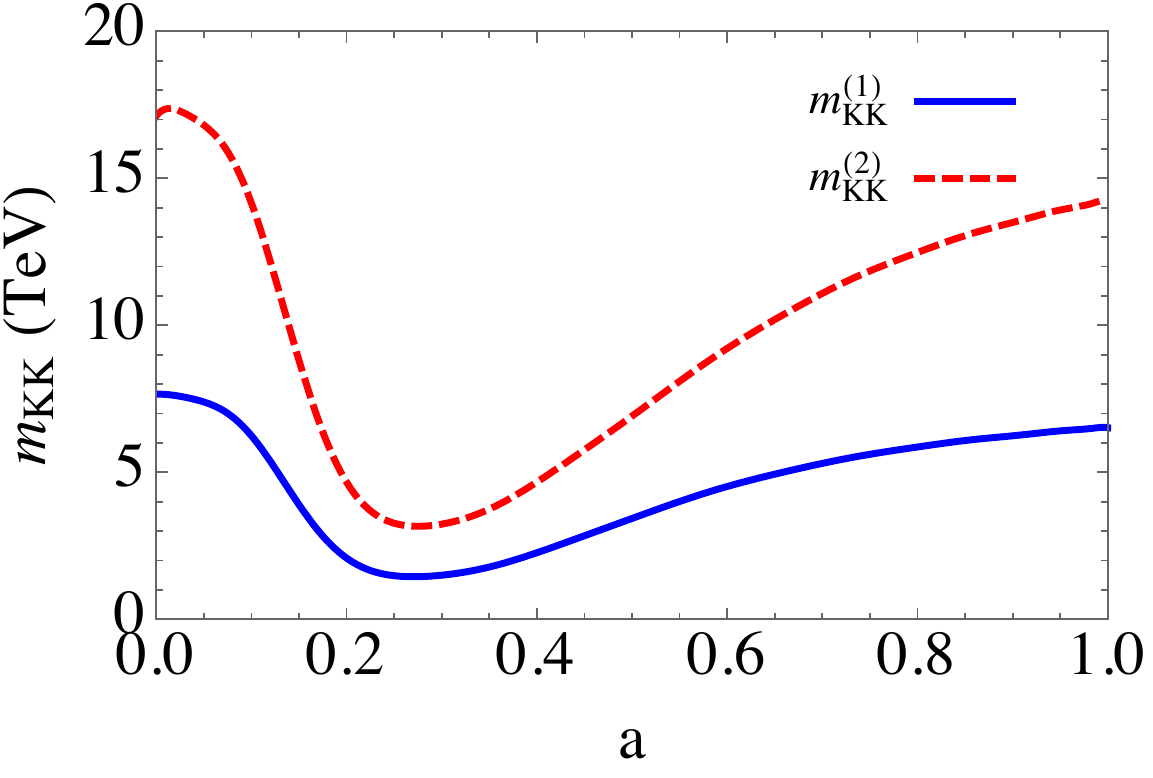}    
\caption{\it Bounds on $m_{KK}$ in TeV, as a function of $a$, for $b=1$, from oblique observables.}
\label{STplot}
\end{center}
\end{figure}
The present experimental bounds on the $S$ and $T$ parameter are given by~\cite{Agashe:2014kda}
\be
T=0.05\pm 0.07,\quad S=0.00\pm 0.08\qquad \textrm{(90\% correlation)}\,.
\ee
These constraints imply a lower bound on the mass of the vector KK modes. The bounds on $m_{KK}$ as a function of $a$
are shown in Fig.~\ref{STplot} (for the choice $b=1$). For different values of $b$ a similar behavior of the bounds is obtained.

As can be seen from the plot in Fig.~\ref{STplot}, for $a \sim 0.25$ a significant reduction in the bound is present.
In particular for the value of the parameters chosen in this paper, $b=1$ and $a=0.2$, the bound on the mass
of the first KK mode is $m_{KK} \gtrsim 2$~TeV.
The reduction in the bounds is one of the peculiar features of soft-wall metrics, as the one we are using in our model.
Even in the absence of a custodial symmetry in the bulk, the contributions to the oblique parameters
are suppressed if the soft wall singularity is not far away from the IR brane.
One way of understanding this effect is to look at the shape of the normalized physical Higgs wave
function in the presence of the soft-wall  metric ($\widetilde h_{SW}$)~\cite{Carmona:2011ib,Quiros:2013yaa}
\be
\widetilde h_{SW}(y) =\frac{\sqrt{y_1}}{\sqrt{\int e^{-2A+2\alpha ky}}}\,e^{-A+\alpha ky}
\ee
It turns out that there is a local maximum of the function $\widetilde h_{SW}$ for values of $y<y_1$,
unlike in the pure RS case in which $\widetilde h_{RS}$ grows monotonically from the UV to the IR brane.
As a consequence $\widetilde h_{SW}(y_1) < \widetilde h_{RS}(y_1)$ and, since the KK-modes are localized towards
the IR brane, there is a significant reduction in the EWSB-induced mass mixing with the gauge zero modes.
This results in a suppression of the corrections to the $S$ and $T$ observables with respect to the RS scenario.

\subsubsection{The $Z\overline b b$ coupling}
\label{Zbbcoupling}

\begin{figure}[htb]
\centering
 \includegraphics[width=12cm]{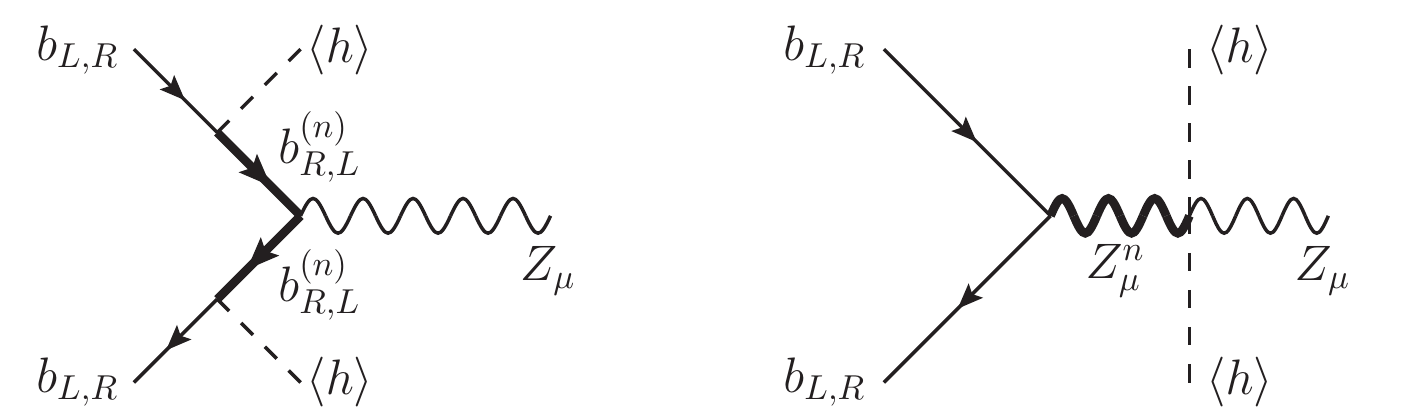} 
\caption{\it Feynman diagrams contributing to  the coupling $Z\bar b b$. }
\label{Zbbdiagram}
\end{figure}

As we mentioned in the general discussion in Sect.~\ref{sec:gen_overview}, the $Z$ boson couplings to SM fermions with a
sizable degree of compositeness can be modified by two independent effects: one coming from the vector KK modes and the
other from the fermion KK excitations. The diagrams induced by these two effects in the case of the
$Z_\mu \bar b_{L,R}\gamma^\mu b_{L,R}$ couplings are shown in Fig.~\ref{Zbbdiagram}.
The distortion in the couplings can be straightforwardly written as a sum over the contributions of the various KK modes,
as we did to derive the approximate result in Eq.~(\ref{eq:delta_Zff}). It is however possible to carry out explicitly the sum over
the KK levels, thus obtaining the full result~\cite{Cabrer:2011qb}
\begin{equation}\label{eq:delta_g}
\delta g_{b_{L,R}}= - g_{b_{L,R}}^{SM}m_Z^2\widehat \alpha_{b_{L,R}}\pm g\frac{v^2}{2}\widehat\beta_{b_{L,R}}\,,
\end{equation}
where $g_{b_{L,R}}^{SM}$ denote the (tree-level) $Z$ coupling to the $b_{L,R}$ fields in the SM, while
\begin{align}
\widehat\alpha_{b_{L,R}}= & y_1\int_0^{y_1} e^{2A}\left(\Omega_h-\frac{y}{y_1}\right)\left(\Omega_{b_{L,R}}-1\right)\,,\nonumber\\
\widehat\beta_{b_{L,R}}= & y_b^2\int_0^{y_1} e^{2A}\left( \frac{d \Omega_{b_{R,L}}}{dy}\right)^{-1}\left(\Gamma_b-\Omega_{b_{R,L}}\right)^2\,,\label{eq:defs1}
\end{align}
with $y_b$ the bottom quark Yukawa coupling and
\be\label{eq:defs2}
\Omega_{b_{L,R}}=\frac{\displaystyle \int_0^y e^{(1-2c_{b_{L,R}})A}}{\displaystyle \int_0^{y_1 }e^{(1-2c_{b_{L,R}})A}}\,,\qquad
\Gamma_b=\frac{\displaystyle \int_0^y he^{-(c_{b_L}+c_{b_R})A}}{\displaystyle \int_0^{y_1} he^{-(c_{b_L}+c_{b_R})A}}\,.
\ee
It is easy to recognize that the two terms in Eq.~(\ref{eq:delta_g}) correspond respectively to the effects of the
massive vectors and of the fermion KK modes~\footnote{In our computation we are not including additional contributions coming
from the mixing of the bottom quark with the light-generations quarks that is induced after EWSB. This effect is however
negligible since the mixing matrix is highly hierarchical.}.

\begin{figure}[t]
\centering
     \includegraphics[width=7.7cm]{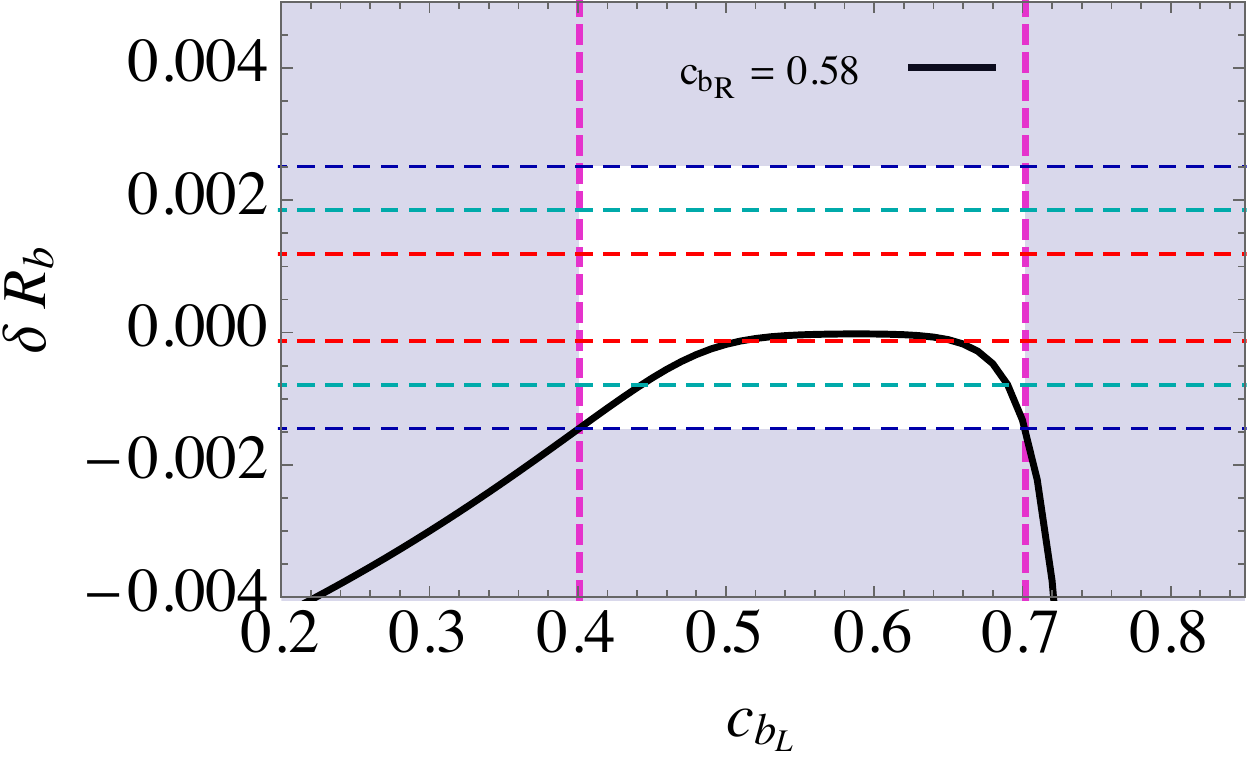}
     \hfill
   \includegraphics[width=7.7cm]{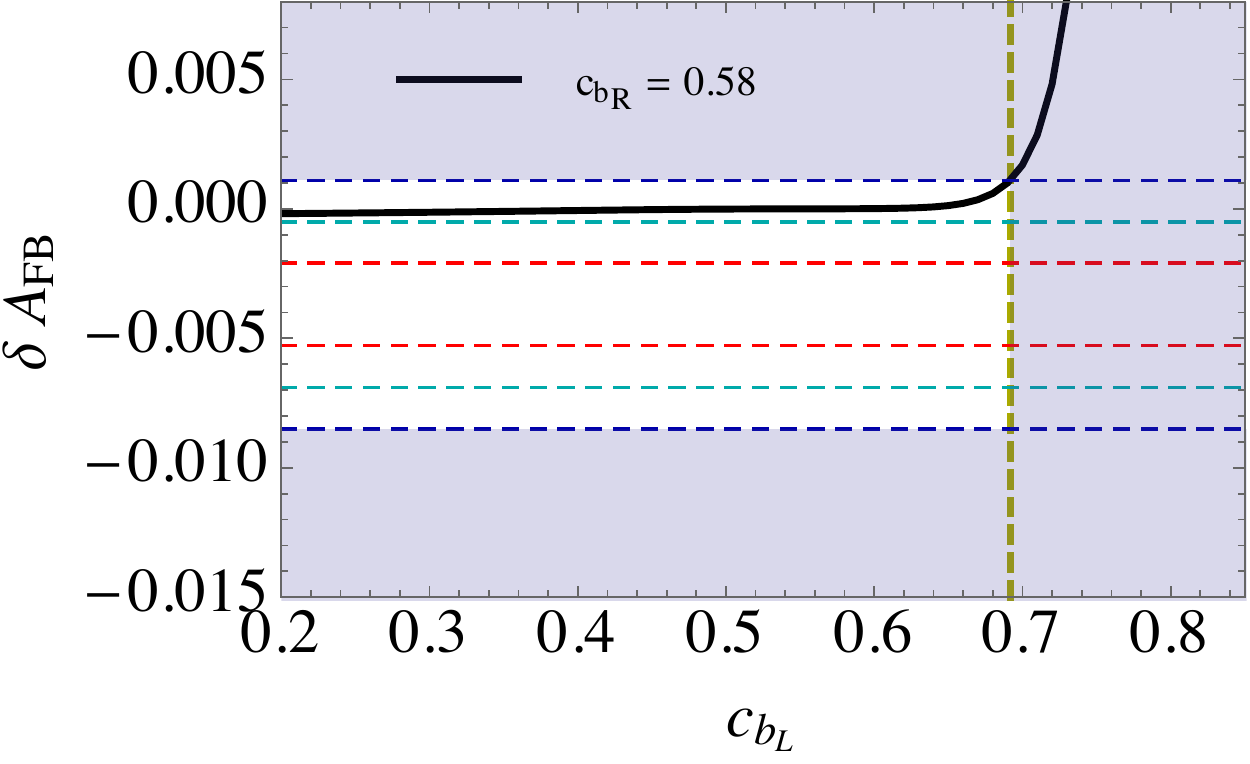}  
 \caption{\it Value of $\delta R_b$ (left panel) and $\delta A_{FB}$ (right panel) as a function of $c_{b_L}$
 obtained in a scenario with no tuning in the Higgs sector ($\alpha = 3$).
 In both panels we fixed $c_{b_R}=0.58$ (other values of $c_{b_R}$ do not lead to significant changes).
 The horizontal lines in both figures correspond to the experimental bounds at $1\sigma$, $2\sigma$ and $3\sigma$.
 }
\label{fig:dRbdAb}
\end{figure}

\begin{figure}[t]
\centering
     \includegraphics[width=7.5cm]{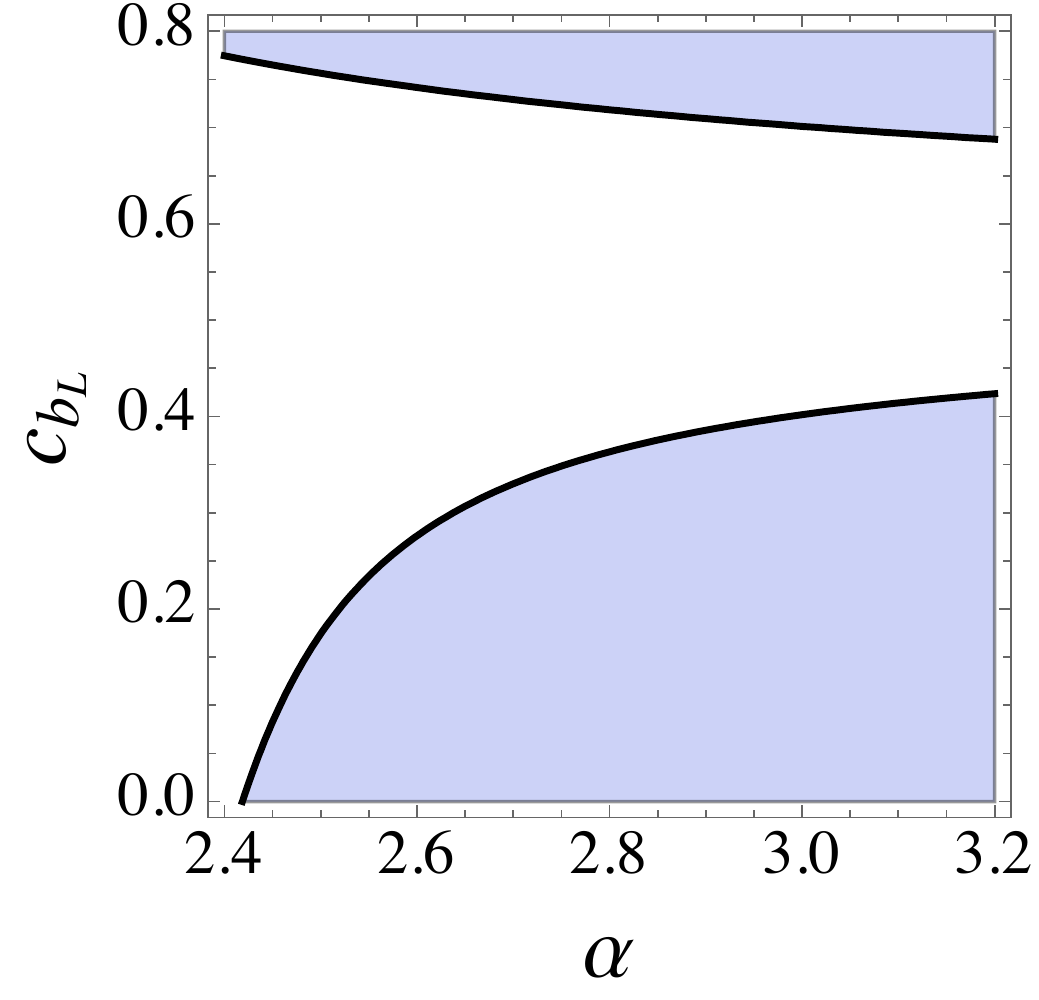}
     \hfill
 \includegraphics[width=7.5cm]{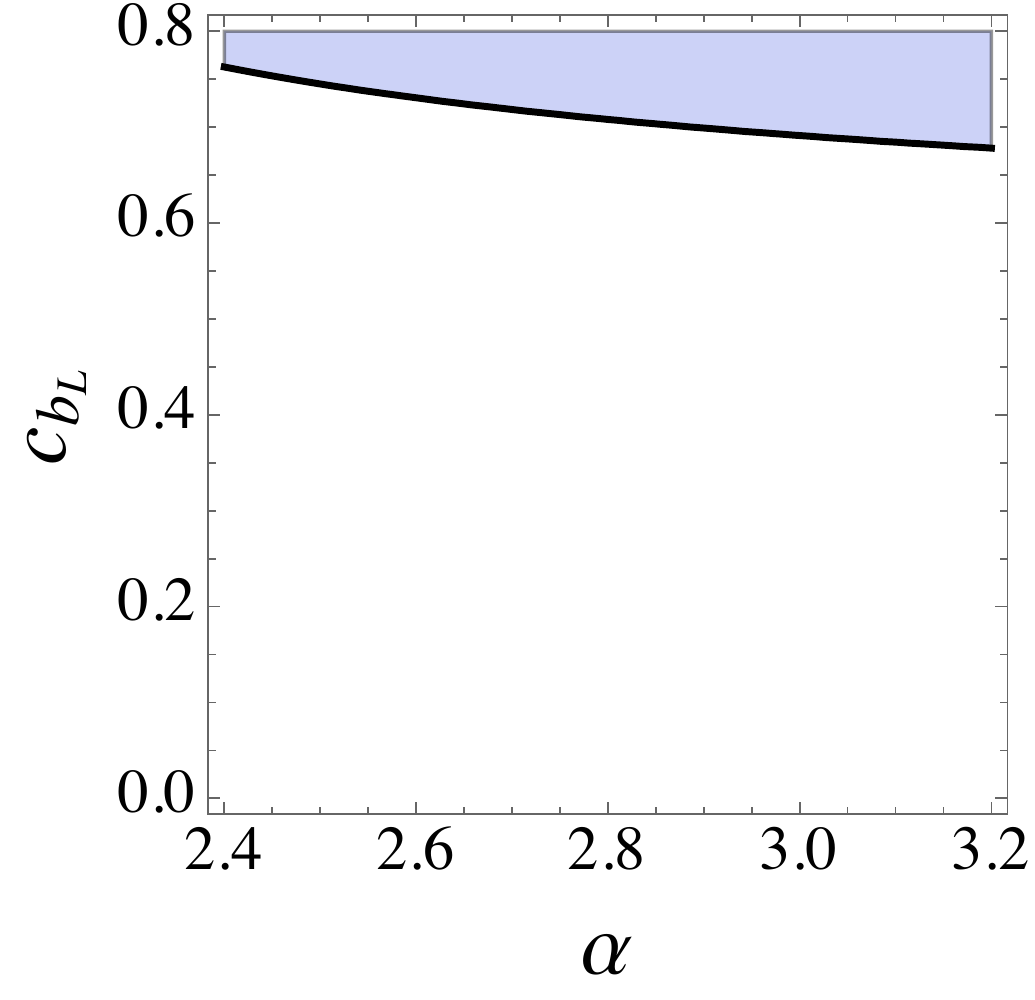}
 \caption{\it Region in the plane $(\alpha,c_{b_L})$ allowed (unshadowed area) by experimental data on $\delta R_b$ (left panel) and $\delta A_{FB}^b$ (right panel) at the $3\sigma$ level. In both panels we have fixed $c_{b_R}=0.58$ (other values of $c_{b_R}$ do not lead to significant changes).
  }
\label{fig:dRbdAb1}
\end{figure}

The experimental constraints on the $Z\overline b b$ coupling come from the $Z$-pole observables measured at LEP,
in particular $R_b$, the ratio of the $Z\to \bar b b$ partial width to the inclusive hadronic width, and $A_{FB}^b$,
the forward-backward asymmetry of the bottom quark. In addition SLD directly measured the bottom asymmetry $A_b$
with longitudinal beam polarizations. These three quantities are related at tree-level to the $Z\bar b b$ couplings by
\be
R_b=\frac{g_{b_L}^2+g_{b_R}^2}{g_{b_L}^2+g_{b_R}^2+\sum_{q\neq t,b}(g_{q_L}^2+g_{q_R}^2)}\,,\quad A_b=\frac{g_{b_L}^2-g_{b_R}^2}{g_{b_L}^2+g_{b_R}^2}\,,\quad
A_{FB}^b=\frac{3}{4}\frac{g_{e_L}^2-g_{e_R}^2}{g_{e_L}^2+e_{b_R}^2}A_b\,.
\ee
From these expressions we can extract the contributions induced by the coupling modifications
\begin{align}
\delta R_b= & -0.606 \,\delta g_{b_L}+0.110\, \delta g_{b_R}\,,\nonumber\\
 \delta A_b= & -0.227\, \delta g_{b_L}-1.25\,\delta g_{b_R}\,,\\ 
 \delta A_{FB}^b= & -0.026\, \delta g_{b_L}-0.141\, \delta g_{b_R}\,.\nonumber
\end{align}
The experimental values for the above observables are given by~\cite{Agashe:2014kda}
\begin{align}
\delta R_b^{exp} & = 0.00053\pm 0.00066\,,\nonumber\\
\delta A_b^{exp} & = -0.0117\pm 0.020\,,\\
\delta A_{FB}^{b,\, exp} & = -0.0037\pm 0.0016\,.\nonumber
\end{align}

The values of $\delta R_b$ and $\delta A_{FB}$ as functions of $c_{b_L}$ are plotted in Fig.~\ref{fig:dRbdAb} (for $c_{b_R} = 0.58$).
To derive these results we chose $\alpha = 3$, which corresponds to a completely Natural scenario with no tuning in the Higgs sector.
One can see that already in the SM ($\delta R_b = \delta A_{FB} = 0$) some tension slightly above the $2\sigma$ level is present
in the data. In our model we are never able to reduce the tension, although for a sizable range of the parameters the agreement
with the data is the same as in the SM. We extract a constraint on $c_{b_L}$ by allowing only configurations which agree with the
data within the $3\sigma$ range. In this way we find $0.4 \lesssim c_{b_L} \lesssim 0.7$.

If some amount of tuning is allowed in the Higgs sector (i.e.~for $\alpha < 3$), the Higgs background wave-function becomes
more flat and the EWSB mixing among different KK levels decreases. In this case the corrections to the $Z$ couplings can be
reduced and the bounds on the $b$ compositeness relaxed. We show in Fig.~\ref{fig:dRbdAb1} how the bounds on $c_{b_{L,R}}$
vary as a function of $\alpha$. The amount of fine tuning as a function of $\alpha$ was plotted in Fig.~\ref{fermionmasses}.
If we allow for a $5\%$ tuning (corresponding to $\alpha \simeq 2.9$), the bounds are slightly relaxed to
$0.37 \lesssim c_{b_L} \lesssim 0.71$. Notice however that the tuning increases exponentially for $\alpha \lesssim 2.95$, thus a further significant reduction of the bounds would require unacceptably high tuning.

\subsubsection{The $Z{\overline \mu}\mu$ coupling}
\label{Zmumucoupling}

Analogously to the bottom couplings to the $Z$, the massive KK modes also induce modifications of the muon couplings. The explicit formulae for these corrections can be obtained from
Eqs.~(\ref{eq:delta_g}), (\ref{eq:defs1}) and (\ref{eq:defs2}) with obvious substitutions.

The current bounds on the distortions of the muon couplings to the $Z$ are given in Ref.~\cite{Agashe:2014kda}.
In our scenario we assumed that the right-handed muon component is almost elementary, thus its coupling
does not deviate appreciably from its SM value, $\delta g_{\mu_{R}}(c_{\mu_{R}})\simeq 0$. In this case the bounds on the deviation of the $\mu_L$ coupling are given by
$\left|\delta g_{\mu_{L}}(c_{\mu_{L}})/g^{SM}_{\mu_{L}}\right|\lesssim 5\times 10^{-3}$ at $95\%$~CL.

\begin{figure}[htb]
\centering
       \includegraphics[width=7.8cm]{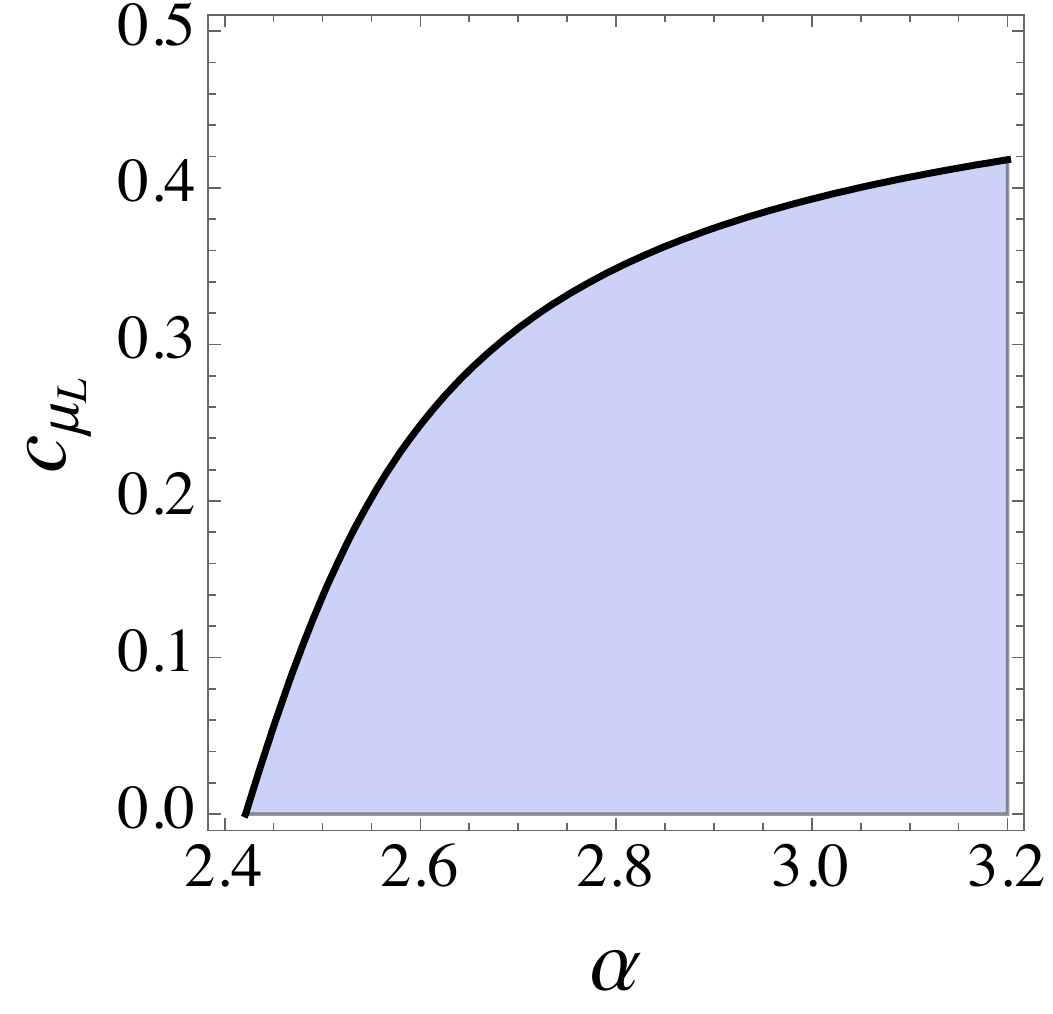}
\caption{\it Region in the plane $(\alpha,c_{\mu_L})$ with the contour line $\delta g_{\mu_{L,R}}^Z(c_{\mu_{L,R}})/g^{Z,\,SM}_{\mu_{L,R}} \leq 5 \times 10^{-3}$. 
}\label{figQ1}
\end{figure}

As in the case of the bottom coupling, also the $\mu_L$ coupling to the $Z$ depends on the amount of compositeness of the
fermion as well as on the localization of the Higgs, i.e.~by the parameter $\alpha$. Smaller values of $\alpha$ imply a reduction of the
corrections, at the price of an increased amount of tuning in the Higgs sector. The bound on $c_{\mu_L}$ as a function
of $\alpha$ is shown in Fig.~\ref{figQ1}.
Requiring our model to be completely Natural implies the constraint $c_{\mu_L} \gtrsim 0.4$. If we accept a tuning of the order of $5\%$ the constraint is
relaxed to $c_{\mu_L} \gtrsim 0.37$.

\subsection{Flavor observables}
\label{flavor}

Another important set of constraints comes from $\Delta F = 2$ flavor-changing processes mediated by four-fermion
contact interactions. These observables can be used to put some bounds on the amount of compositeness of the bottom quark.

As we already discussed in Sect.~\ref{sec:gen_overview}, the main new-physics contributions to $\Delta F = 2$ processes come
from the exchange of gluon KK modes. The leading flavor-violating couplings of the KK gluons involving the down-type quarks are given by (compare with Eq.~(\ref{lagrangianEW_od}))
\begin{align}
\mathcal L_s = G_{n\mu}^A\Big[
V_{3i}^* V_{3j}\,\bar d_i\gamma^\mu \left\{\left(g^{G^n}_{b_{L}}-\overline g^{G^n}_{L}  \right)P_L+\left(g^{G^n}_{b_R}-\overline g^{G^n}_{R}  \right)P_R\right\} d_j + {\rm h.c.} \Big]\,,\label{eq:lagrangian_QCD}
\end{align}
where the $g^{G^n}_{f_{L,R}}$ couplings are given by
\be
g^{G^n}_{f_{L,R}} =  g_s f^{(n)}(c_{f_{L,R}})\,.
\ee
In Eq.~(\ref{eq:lagrangian_QCD}), $\overline g^{G^n}_{L,R}$ denote the KK-gluon couplings to the first and second generation down quarks,
which are universal due to the $U(2)$ flavor symmetry present in our model, namely
$\overline g^{G^n}_{L,R} = \overline g^{G^n}_{d_{1_{L,R}}} = \overline g^{G^n}_{d_{2_{L,R}}}$.

After integrating out the massive KK gluons, the couplings in Eq.~(\ref{eq:lagrangian_QCD}) give rise to the following set of $\Delta F=2$ dimension-six operators
\begin{align}
{\cal L}_{\Delta F = 2} = \sum_n \Bigg\{&\frac{c_{ij}^{LL(n)}}{M_n^2} (\overline d_{iL} \gamma^\mu d_{jL}) (\overline d_{iL} \gamma_\mu d_{jL})
+ \frac{c_{ij}^{RR(n)}}{M_n^2} (\overline d_{iR} \gamma^\mu d_{jR}) (\overline d_{iR} \gamma_\mu d_{jR})\nonumber\\
&+ \frac{c_{ij}^{LR(n)}}{M_n^2} (\overline d_{iR} d_{jL}) (\overline d_{iL} d_{jR})\Bigg\}\,,
\end{align}
where 
\begin{align}
c^{LL,RR(n)}_{ij} & = \frac{1}{6}\left(V_{3i}^\star V_{3j}\right)^2\left(g_{b_{L,R}}^{G^n}-\overline g_{L,R}^{G^n}\right)^2\,,\nonumber\\
\rule{0pt}{1.5em} c^{LR(n)}_{ij} & = \left(V_{3i}^\star V_{3j}\right)^2\left(g_{b_{L}}^{G^n}-\overline g_{L}^{G^n}\right)
\left(g_{b_{R}}^{G^n}-\overline g_{R}^{G^n}\right)\,.
\end{align}
The current bounds on the $\Delta F = 2$ contact operators~\cite{UTfit,Isidori:2013ez}
can be translated into constraints on the quantities
\begin{align}
\sum_n \frac{\left(g_{b_{L,R}}^{G^n}-\overline g_{L,R}^{G^n}  \right)^2}{M_n^2[{\rm TeV}]} &
\simeq \sum_n \frac{\left(g_{b_{L,R}}^{G^n}\right)^2}{M_n^2[{\rm TeV}]} \leq 0.14 \,, \label{eq:bound1} \\
\sum_n \frac{\left(g_{b_L}^{G^n}- \overline g_{L}^{G^n}  \right)\left(g_{b_R}^{G^n}-\overline g_{R}^{G^n}  \right)}{M_n^2[{\rm TeV}]}
& \simeq \sum_n \frac{g_{b_L}^{G^n} g_{b_R}^{G^n}}{M_n^2[{\rm TeV}]} \leq 3\times10^{-4} \,, \label{boundBs}
\end{align}
where, to simplify the full expressions, we used the fact that $\overline g_{L,R}^{G^n} \ll g_{b_{L,R}}^{G^n}$
which follows from our assumption that the light-generation quarks are almost elementary.

\begin{figure}[htb]
  \centering
   \includegraphics[width=7.8cm]{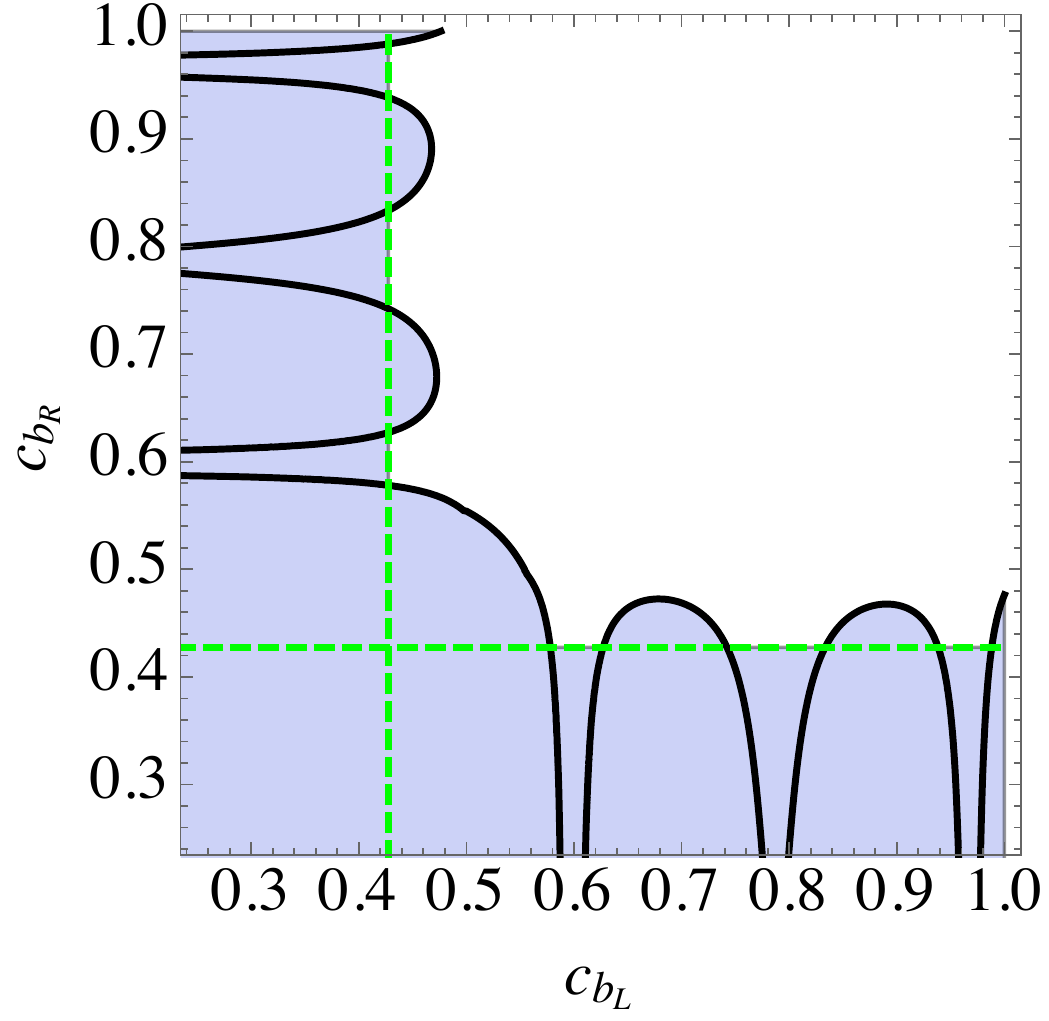}
\caption{\it Region in the plane $(c_{b_L},c_{b_R})$ that accommodates the bound of Eqs.~(\ref{eq:bound1}) and~(\ref{boundBs}). The dashed green lines
represent the constraints in Eq.~(\ref{cbL}). The allowed points correspond to the unshaded region.
}\label{fig:flavor}
\end{figure}

The bound in Eq.~(\ref{eq:bound1}) is quite interesting since it allows to derive independent constraints on the amount of compositeness
of the $b_L$ and $b_R$ components or, in other words, on the $c_{b_L}$ and $c_{b_R}$ parameters.
At the $95\%$~CL we find
\be
c_{b_{L,R}} \geq 0.43\,.
\label{cbL}
\ee
The constraint in Eq.~(\ref{boundBs}) gives instead a combined bound on the $b_L$ and $b_R$ compositeness.
The allowed configurations in the $(c_{b_L}, c_{b_R})$ plane are shown in Fig.~\ref{fig:flavor}. One can see that
the bound from the LR operators dominates for $c_{b_L} \simeq c_{b_R} \simeq 0.5$, whereas if only one bottom chirality has a sizable compositeness, the bounds from LL, RR and LR operators are of the same order.

\subsection{The ATLAS di-muon resonance search}
\label{dimuon}

An additional experimental constraint comes from direct searches for high-mass resonances decaying
into di-muon final states. This search has been performed by the ATLAS Collaboration
both at $\sqrt{s}=8$~TeV with an integrated luminosity of 20.3~fb$^{-1}$~\cite{Aad:2014cka},
and at $\sqrt{s}=13$~TeV with an integrated luminosity of 3.2~fb$^{-1}$~\cite{ATLASdimuon}.

\begin{figure}[ht]
\centering
 \includegraphics[width=8cm]{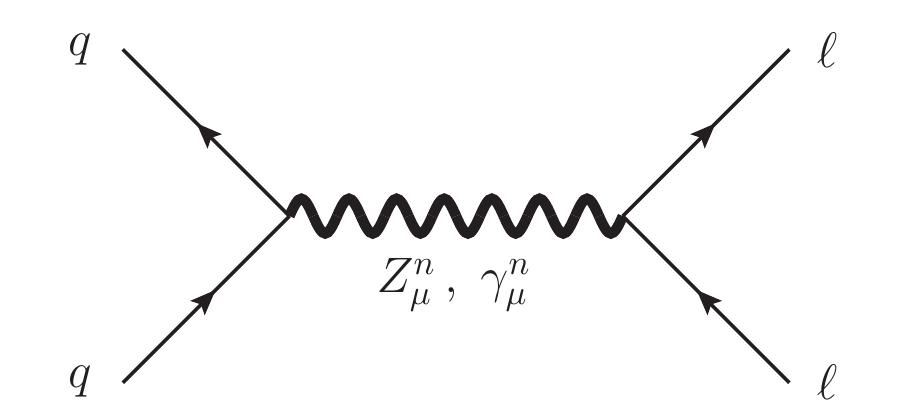}  
\caption{\it Drell-Yan contribution to the process $\sigma(pp\to X_n\to\ell\ell)$. }
\label{DY}
\end{figure}

In our model this search can be sensitive to the production of the massive excitations of the $Z$ and of the photon. The resonances $Z^n_\mu$ and $\gamma^n_\mu$ can be produced by Drell-Yan processes and decay into a pair of leptons as shown in the diagram of Fig.~\ref{DY}.
The decay width of $X^n$ ($X=Z,\gamma$) into a fermion pair $\bar f f$ is given at the tree-level by
\be
\Gamma(X_n\to f\bar f)=N_cM_n\frac{\sqrt{2}G_Fm_Z^2}{24\pi g^2}\left[\left(g_{f_L}^{X_n} \right)^2+\left(g_{f_R}^{X_n} \right)^2\right]\,,
\ee 
where $N_c=3$ ($1$) for quarks (leptons) and we neglected the effect of fermion masses.
In the narrow width approximation (NWA) the cross-section for the process $pp\to Z^n/\gamma^n \to \mu^+\mu^-$
approximately scales as
\begin{align}
&\sigma(pp\to Z_n/\gamma_n\to \mu^+\mu^-)\propto A=\sum_{X=Z,\gamma}A_{X^n}\,,\nonumber\\
A_{X^n}&=\frac{g_{\mu_L}^2\left(2g_{u_L}^2+2g_{u_R}^2+g_{d_L}^2+g_{d_R}^2\right)}
{3(g^2_{b_L} + g^2_{t_L} + g^2_{b_R} + g^2_{t_R}) + \sum_{\ell = \mu, \tau} (g^2_{\ell_L}+g^2_{\nu_\ell}+g^2_{\ell_R})}\,,\label{brappmu}
\end{align}
where all couplings refer to the $g_{f_{L,R}}^{X^n}$ couplings and for simplicity we have omitted the superscript $X^n$. To obtain the above formula we
neglected, in the denominator of Eq.~(\ref{brappmu}), the contributions from the light quarks ($u, d, s, c$) and leptons ($e, \nu_e$) to the decay width.

\begin{figure}[htb]
  \centering
  \includegraphics[width=8.cm]{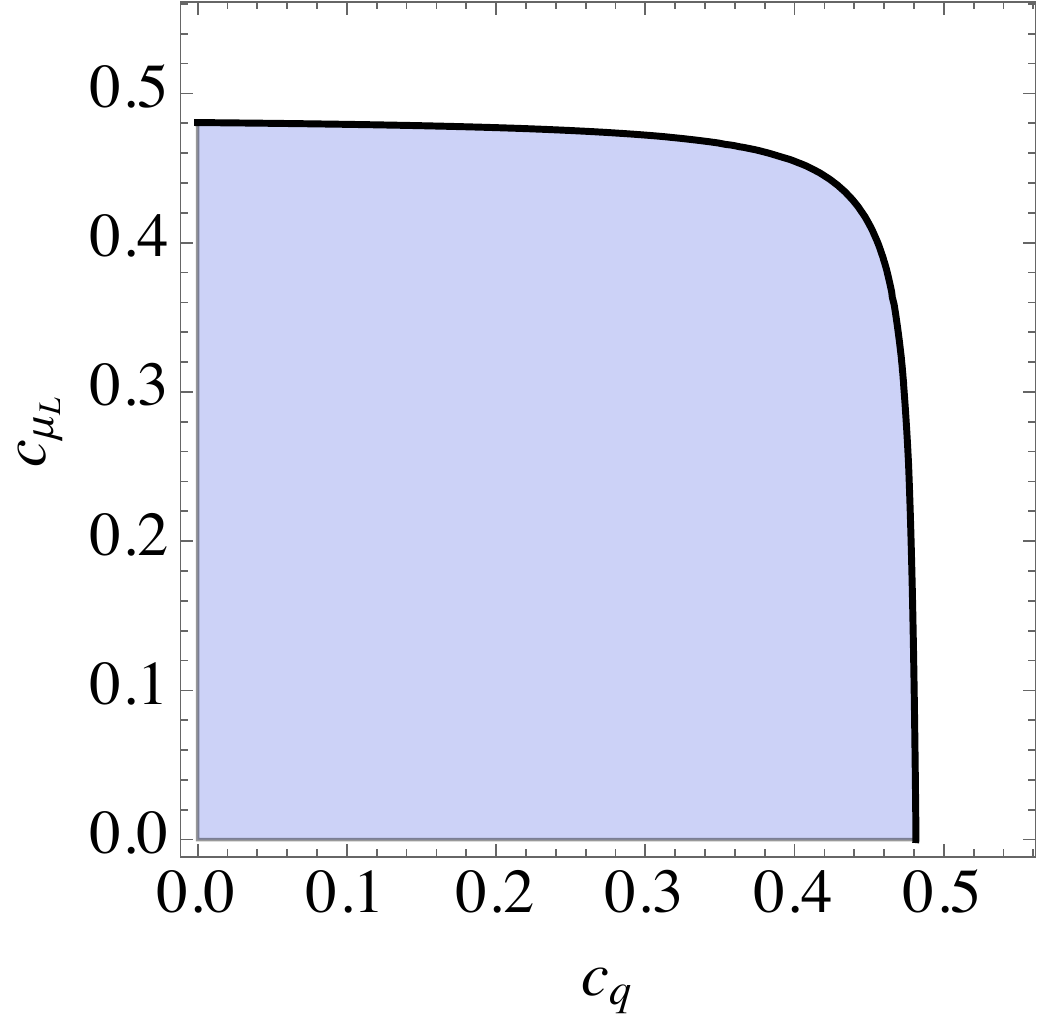}  
\caption{\it Exclusions in the plane $(c_q,c_{\mu_L})$ coming from the searches of massive resonances decaying into di-muons.
The results are derived for the following benchmark choice of the parameters: $c_{\mu_R} = 0.7$, $c_{\tau_R} = 0.6$,
$c_{b_R} = 0.6$, $c_{t_R} = -0.5$, $c_{b_L} = 0.43$, $c_{\tau_L} = c_{\mu_L}$. The remaining parameters have been fixed
by requiring the quark masses to be reproduced for values of the Yukawa couplings $\sqrt{k} \hat Y = 3$.}\label{plotcqcmuL}
\end{figure}

The $13$~TeV ATLAS experimental bound~\cite{ATLASdimuon}, 
translates into the constraint $A<0.022$ ($A<0.003$) for $M_n=3$ TeV ($M_n=2$ TeV) at $95\%$~CL. 
To get an idea of the bounds on the parameter space of our model we reduce the number of free parameters by setting a common value for the compositeness of the light quarks $c_{q_L} = c_{q_R} \equiv c_q$ for $q = u, d$.
The constraints in the plane $(c_q, c_{\mu_L})$ are shown in Fig.~\ref{plotcqcmuL}, for a benchmark choice of the parameters.
The numerical results show that for almost elementary light quarks $c_q \gtrsim 0.46$ basically no bound is present on the amount of $\mu_L$ compositeness.

\subsection{Direct searches on KK gluons}
\label{directgluons}

A second set of direct constraints comes from searches of massive gluon KK modes. Since the gluons are the most strongly coupled KK modes, they should be more copiously produced than EW KK modes and easily detected at the LHC due to the sizable branching ratio into $\bar tt$. On the other hand, since the EWSB contribution to the mass of the KK modes is negligible, it turns out, as was already noticed, that all gauge boson KK modes are approximately degenerate in mass. Therefore the bounds on the mass of gluon KK modes directly translates into bounds on the mass $M_n$ of all vector KK modes.

Direct searches for KK gluons $G^n_\mu$, in the RS scenarios, have been performed by CMS~\cite{Chatrchyan:2013lca}, for an integrated luminosity of $19.7$~fb$^{-1}$ and $\sqrt{s}=8$~TeV, and by ATLAS~\cite{Aad:2015fna}, for an integrated luminosity of $20.3$~fb$^{-1}$ and  $\sqrt{s}=8$~TeV, by measuring the cross-section $\sigma(pp\to \bar tt)$. The KK gluon is produced by the DY mechanism from light quarks and decays preferentially into top quarks. The experimental searches focus on the lepton plus jets final state, where the top pair decays into $W^+bW^-\bar b$, one $W$ decaying leptonically and the other hadronically. The
$95\%$~CL bounds obtained in these searches are
\be
M_1^{RS}\gtrsim 2.5 \textrm{ TeV (CMS)},\qquad M_1^{RS}\gtrsim 2.2 \textrm{ TeV (ATLAS)}\,.
\label{boundsgluons}
\ee
These bounds are derived in a RS set-up with light fermions localized towards the UV brane ($c_{f}>0.5$) whose
coupling to the first KK gluon excitation has the value $g_{G^1_{RS}\bar f f}\simeq 0.2\, g_s$~\cite{Lillie:2007yh}.
In our model, instead, this coupling is significantly smaller, $g_{G^1\bar f f}\simeq 0.13\, g_s$ (see Fig.~\ref{KKcouplings}).
This translates into a slightly weaker bound, namely $M_1\gtrsim 2$~TeV, which coincides with our benchmark choice
for the KK scale.
We want to stress here that our choice of $M_1\simeq 2$~TeV is just a benchmark point and the results found in our analysis
would be essentially unchanged by slightly increasing this value.

\section{Conclusions}
\label{conclusions}

In this paper we explored a modified RS model as a possible explanation of the recently-found anomalies in the semi-leptonic
$B$-meson decays. The attractiveness of our scenario is the fact that the new dynamics that generates the flavor anomalies
is \textit{not ad-hoc}, but instead is intrinsically linked to the mechanism that in RS scenarios ensures a natural solution of the EW
Hierarchy Problem. The corrections to the $B$-meson physics are indeed due to $\Delta F = 1$ four-fermion contact interactions
induced by the exchange of the heavy KK modes of the $Z$-gauge boson and of the photon. These modes are unavoidably
present in the RS scenarios in which the SM gauge invariance is also extended into the bulk.

Sizable corrections to the $\Delta F = 1$ processes can be obtained provided that the left-handed components of the bottom
quark and of the muon lepton have a sizable degree of compositeness, i.e.~that they are sufficiently localized towards the IR brane.
In this case large couplings to the gauge KK modes are generated, which translate into sizable contributions to the
contact operators ${\cal O}_{9, 10}$. Additional contributions to operators involving the right-handed quarks ${\cal O}'_{9, 10}$ could also be generated if the $b_R$ quark is sufficiently IR localized.

The main quantities that control the $\Delta F = 1$ effects are thus the 5D bulk masses of the bottom and the muon fields,
which are conveniently encoded in the $c_{b_{L,R}}$ and $c_{\mu_L}$ parameters. The parameter space region that
allows to fit the current flavor anomalies is shown in Fig.~\ref{LHCb}. The strongest constraints on the parameter space
of our model come from the requirement of inducing a large enough contribution to the ${\cal O}_9$ operator (see Tab.~\ref{fit}).
This implies that at least one of the conditions $c_{\mu_L} \lesssim 0.45$ and $c_{b_L} \lesssim 0.45$
must be satisfied. The $b_R$ field can instead be almost elementary ($c_{b_R} \gtrsim 0.5$).

As a consequence of the large amount of compositeness for the $\mu_L$ and/or $b_L$ fields, sizable corrections in the electroweak and flavor observables can be generated.
Deviations in the $g_{Z\bar\mu\mu}$ and $g_{Z\bar b b}$ couplings are indeed
expected and can be used to put some bounds on $c_{\mu_L}$ and $c_{b_L}$. In a completely Natural model
one finds $c_{\mu_L} \gtrsim 0.4$ and $c_{b_L} \gtrsim 0.4$. These bounds can be slightly relaxed, at the price of introducing
some fine-tuning in the Higgs mass, by modifying the localization of the Higgs field. A fine-tuning of the order of $5\%$
allows to weaken the bounds to $c_{\mu_L}, c_{b_L} \gtrsim 0.37$ (see Figs.~\ref{fig:dRbdAb} and~\ref{figQ1}). Notice however that
the amount of tuning increases exponentially when one tries to further lower these bounds, thus quickly reaching unacceptably high values.

Even stronger bounds can be derived from flavor observables, most noticeably $\Delta F = 2$ processes.
The exchange of massive vector fields (in particular the KK modes of the gluons) gives rise to four-fermion contact interactions
involving the SM quarks. After EWSB these operators can develop flavor-changing components due to the mixing among
the various generations induced by the Yukawa couplings. Contact operators involving the down-type quarks
are fully controlled by the amount of compositeness of the $b_L$ and $b_R$ fields and by the CKM matrix elements. The flavor
data can thus be directly translated into bounds $c_{b_L} \gtrsim 0.43$, and $c_{b_R} \gtrsim 0.43$ which can not
be evaded in our set-up. Even stronger bounds are found for $c_{b_L} \simeq c_{b_R}$, in which case
$c_{b_{L,R}} \gtrsim 0.5$ (see Fig.~\ref{fig:flavor}).

Finally additional constraints on the amount of compositeness of the first generation quarks can be derived from the LHC searches
for resonances decaying into a muon pair, as well as from direct searches of KK gluons. These constraints are however easily fulfilled in our model
by assuming that the first generation quarks are localized toward the UV and thus weakly coupled to the gauge KK modes (see Fig.~\ref{plotcqcmuL}).

\begin{figure}[htb]
\centering
    \includegraphics[width=7.7cm]{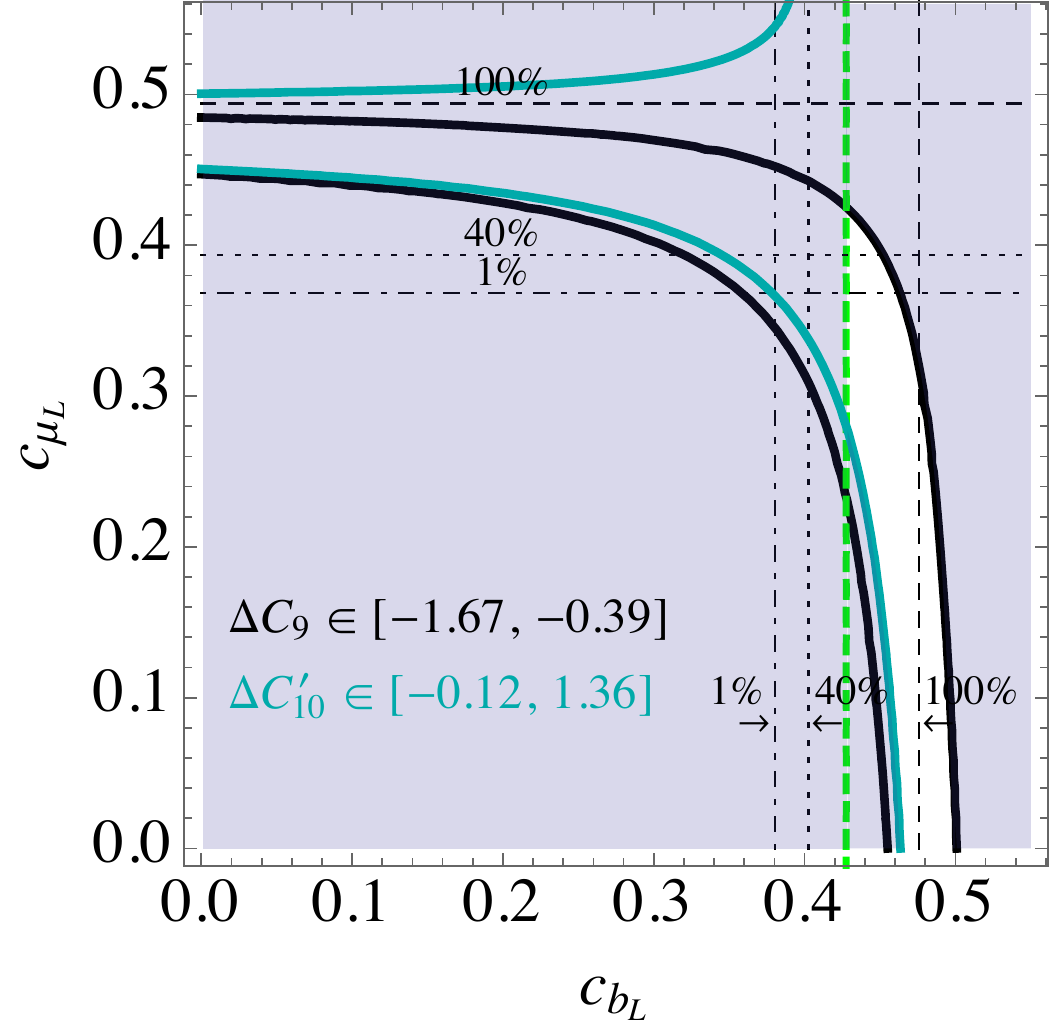} 
    \hfill
  \includegraphics[width=8.cm]{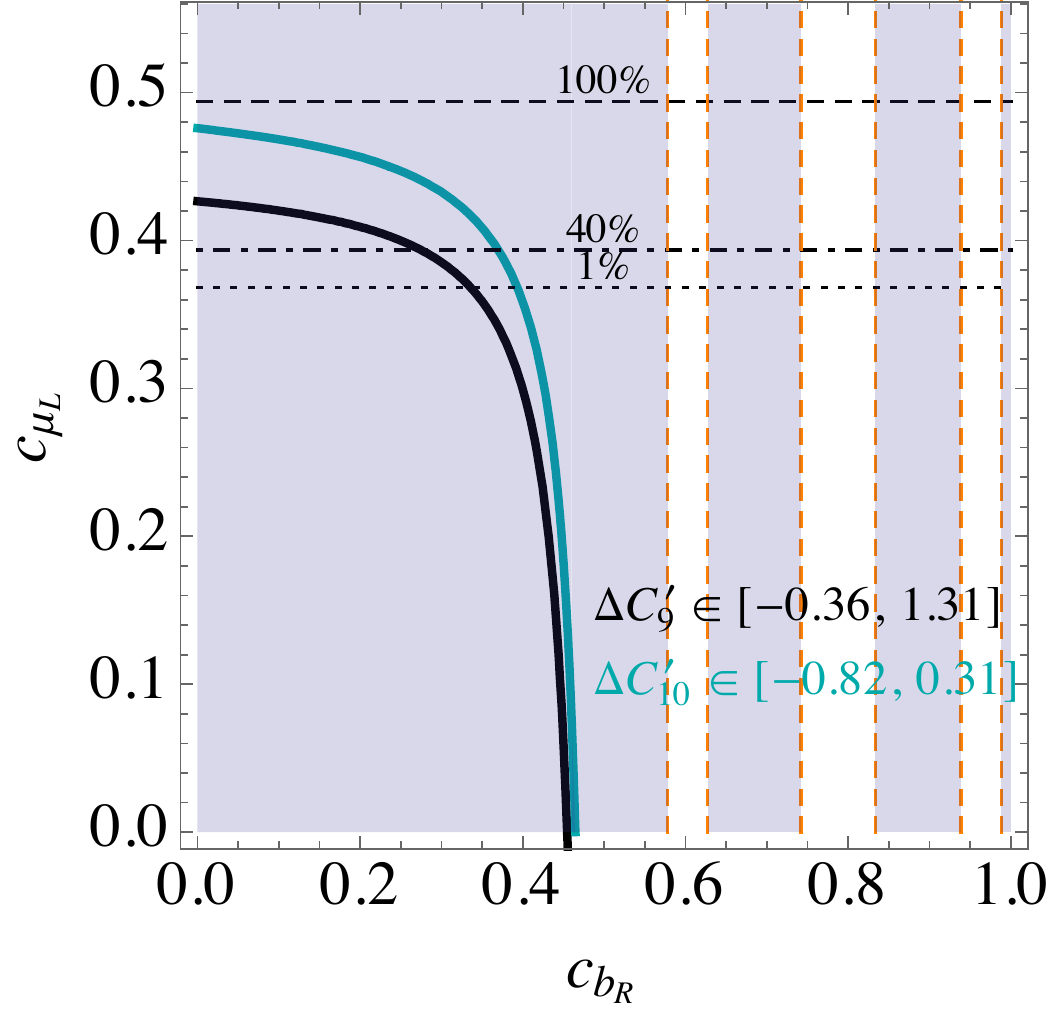}    
\caption{\it Left panel: Region in the plane $(c_{b_L},c_{\mu_L})$ that accommodates $\Delta C_9$ (the band inside the black solid lines) and $\Delta C_{10}$ (the band inside the green solid lines).
The points to the left of the vertical dashed green line are excluded by the flavor bound in Eq.~(\ref{boundBs}).
The fine-tuning needed to pass the constraints on the modification of the $Z\mu\bar \mu$ ($Z\bar bb$) coupling
is shown by the black dashed, dotted and dot-dashed horizontal (vertical) lines (see Sects.~\ref{Zmumucoupling} and \ref{Zbbcoupling}).
Right panel: Region in the plane $(c_{b_R},c_{\mu_L})$  that accommodates $\Delta C_{9}^\prime$ (the region outside the black solid line) and $\Delta C_{10}^\prime$ (the region outside the blue solid line) for the benchmark choice $c_{b_L}=0.43$. The allowed white bands are determined by the flavor constraints.
}
\label{LHCbfinal}
\end{figure}
The results of our analysis are summarized in Fig.~\ref{LHCbfinal}, where we show the parameter space regions
in the $(c_{b_L}, c_{\mu_L})$ and $(c_{b_R}, c_{\mu_L})$ planes that allow to fit the flavor anomalies, together with the
constraints coming from EW and flavor precision measurements.
The regions allowed by all constraints correspond to the un-shadowed areas. Notice that in the plots we did not include the
lower bound for $c_{\mu_L}$ implied by the corrections to the $Z\overline \mu_L \mu_L$ vertex. We instead showed
the amount of fine tuning in the Higgs sector that is needed to pass the EW constraints for each value of $c_{\mu_L}$.
The dashed horizontal line corresponds to a completely Natural scenario, whereas the dotted and dot-dashed lines correspond respectively to $40\%$ and $1\%$ tuning. Analogously, the constraints from the corrections to the $Z\bar bb$ couplings (in particular the observables $\delta A_b$ and $\delta A^b_{FB}$) corresponding to a certain level of tuning are shown by the
dashed, dotted and dot-dashed vertical lines in the left plot. These constraints are however weaker than the one
derived from flavor observables, $c_{b_L} > 0.43$ (shown by the dashed green line in the plot).

To obtain the plot in the right panel we fixed $c_{b_L}=0.43$, the minimum value allowed by $\Delta F=2$ flavor constraints.
For larger value of $c_{b_L}$, i.e.~$c_{b_L} \gtrsim 0.45$, the allowed region in the right panel
plot would be $c_{b_R}\gtrsim 0.55$ as one can infer from Fig.~\ref{fig:flavor}.

An interesting outcome of our analysis is the fact that, in a completely Natural model with no tuning in the Higgs sector,
requiring the flavor anomalies to be reproduced almost completely fix the values of the parameters
$c_{b_L} \simeq c_{\mu_L} \simeq 0.45$. In this configuration the corrections to the $\Delta F = 2$ observables
and the deviations in the $Z$ couplings to the $b_L$ and $\mu_L$ fields are all expected to be close to the present experimental
bounds. This is a quite sharp prediction of our scenario, which predicts correlated deviations potentially testable in not so far future experiments.

Finally, let us also add that the concrete model studied here also sharply predicts the presence of a dilaton-like scalar with mass around $100$~GeV, see Ref.~\cite{Megias:2015ory}. As discussed in Refs.~\cite{CPR,Bellazzini:2013fga,Coradeschi:2013gda,Megias:2014iwa} such a light dilaton can appear naturally in this kind of models, and it is experimentally allowed basically because its couplings to SM fields are slightly suppressed~\cite{Megias:2015ory,Megias:2015qqh}. This dilaton-like state looks quite unrelated to the flavor physics and there might be other extra-dimensional models that manage to address the flavor anomalies and Naturalness without it. In the class of models analyzed here, its presence is related to the fact that the model passes all tests with a low KK scale $\sim 2$~TeV.

\section*{\sc Acknowledgments}

The work of G.P., O.P. and M.Q.~is partly supported by MINECO under Grant
CICYT-FEDER-FPA2014-55613-P, by the Severo Ochoa Excellence Program of
MINECO under the grant SO-2012-0234 and by Secretaria d'Universitats i
Recerca del Departament d'Economia i Coneixement de la Generalitat de
Catalunya under Grant 2014 SGR 1450.  E.M. would like to thank the
Institut de F\'{\i}sica d'Altes Energies (IFAE), Barcelona, Spain,
and the Instituto de F\'{\i}sica Te\'orica CSIC/UAM,
  Madrid, Spain, for their hospitality during the completion of the
final stages of this work. The work of E.M. is supported by the
European Union under a Marie Curie Intra-European Fellowship
(FP7-PEOPLE-2013-IEF) with project number PIEF-GA-2013-623006.


\begin{thebibliography}{99}


\bibitem{Aaij:2014ora}
  R.~Aaij {\it et al.} [LHCb Collaboration],
  Phys.\ Rev.\ Lett.\  {\bf 113} (2014) 151601
  [\hhref{1406.6482} [hep-ex]].


\bibitem{Aaij:2015oid}
  R.~Aaij {\it et al.} [LHCb Collaboration],
  JHEP {\bf 1602} (2016) 104
  [\hhref{1512.04442} [hep-ex]].

\bibitem{Abdesselam:2016llu}
  A.~Abdesselam {\it et al.} [Belle Collaboration],
  \hhref{1604.04042} [hep-ex].
  
\bibitem{Altmannshofer:2013foa}

  S.~Descotes-Genon, J.~Matias and J.~Virto,
  Phys.\ Rev.\ D {\bf 88}, 074002 (2013)
  doi:10.1103/PhysRevD.88.074002
  [\hhref{1307.5683} [hep-ph]]; 
%
  W.~Altmannshofer and D.~M.~Straub,
  Eur.\ Phys.\ J.\ C {\bf 73} (2013) 2646
  [\hhref{1308.1501} [hep-ph]];
%
  R.~Gauld, F.~Goertz and U.~Haisch,
  Phys.\ Rev.\ D {\bf 89} (2014) 015005
  [\hhref{1308.1959} [hep-ph]];
%
  W.~Altmannshofer, S.~Gori, M.~Pospelov and I.~Yavin,
  Phys.\ Rev.\ D {\bf 89} (2014) 095033
  [\hhref{1403.1269} [hep-ph]];
   A.~Crivellin, G.~D'Ambrosio and J.~Heeck,
  Phys.\ Rev.\ Lett.\  {\bf 114}, 151801 (2015)
  [\hhref{1501.00993} [hep-ph]];
%
  D.~Aristizabal Sierra, F.~Staub and A.~Vicente,
  Phys.\ Rev.\ D {\bf 92} (2015) no.1,  015001
  [\hhref{1503.06077} [hep-ph]];
%
  A.~Crivellin, L.~Hofer, J.~Matias, U.~Nierste, S.~Pokorski and J.~Rosiek,
  Phys.\ Rev.\ D {\bf 92} (2015) no.5,  054013
  [\hhref{1504.07928} [hep-ph]];
%
  A.~Celis, J.~Fuentes-Martin, M.~Jung and H.~Serodio,
  Phys.\ Rev.\ D {\bf 92} (2015) no.1,  015007
  [\hhref{1505.03079} [hep-ph]];
%
 W.~Altmannshofer and I.~Yavin,
  Phys.\ Rev.\ D {\bf 92} (2015) no.7,  075022
  [\hhref{1508.07009} [hep-ph]];
  A.~Falkowski, M.~Nardecchia and R.~Ziegler,
  JHEP {\bf 1511} (2015) 173
  [\hhref{1509.01249} [hep-ph]];
  %
  S.~Descotes-Genon, L.~Hofer, J.~Matias and J.~Virto,
  JHEP {\bf 1606} (2016) 092
  [\hhref{1510.04239} [hep-ph]];
%
  B.~Allanach, F.~S.~Queiroz, A.~Strumia and S.~Sun,
  Phys.\ Rev.\ D {\bf 93} (2016) no.5,  055045
  [\hhref{1511.07447} [hep-ph]];
 %
  S.~M.~Boucenna, A.~Celis, J.~Fuentes-Martin, A.~Vicente and J.~Virto,
  Phys.\ Lett.\ B {\bf 760}, 214 (2016)
  doi:10.1016/j.physletb.2016.06.067
  [\hhref{1604.03088} [hep-ph]];
  %
    D.~Buttazzo, A.~Greljo, G.~Isidori and D.~Marzocca,
    \hhref{1604.03940} [hep-ph];
  S.~M.~Boucenna, A.~Celis, J.~Fuentes-Martin, A.~Vicente and J.~Virto,
  JHEP {\bf 1612}, 059 (2016)
  doi:10.1007/JHEP12(2016)059
  [\hhref{1608.01349} [hep-ph]].

\bibitem{Niehoff:2015bfa}
  C.~Niehoff, P.~Stangl and D.~M.~Straub,
  Phys.\ Lett.\ B {\bf 747} (2015) 182
  [\hhref{1503.03865} [hep-ph]].

  
\bibitem{Descotes-Genon:2016hem}
  S.~Descotes-Genon, L.~Hofer, J.~Matias and J.~Virto,
  JHEP {\bf 1606}, 092 (2016)
  doi:10.1007/JHEP06(2016)092
  [\hhref{1510.04239} [hep-ph]];
%
  S.~Descotes-Genon, L.~Hofer, J.~Matias and J.~Virto,
  \hhref{1605.06059} [hep-ph].
  
  
\bibitem{Kosnik:2012dj}
  N.~Kosnik,
  Phys.\ Rev.\ D {\bf 86} (2012) 055004
  [\hhref{1206.2970} [hep-ph]];
  %
  G.~Hiller and M.~Schmaltz,
  Phys.\ Rev.\ D {\bf 90} (2014) 054014
  [\hhref{1408.1627} [hep-ph]];
  %
  B.~Gripaios, M.~Nardecchia and S.~A.~Renner,
  JHEP {\bf 1505} (2015) 006
  [\hhref{1412.1791} [hep-ph]];
  %
  S.~Sahoo and R.~Mohanta,
  Phys.\ Rev.\ D {\bf 91} (2015) no.9,  094019
  [\hhref{1501.05193} [hep-ph]];
  %
  D.~Becirevic, S.~Fajfer and N.~Ko¨nik,
  Phys.\ Rev.\ D {\bf 92} (2015) no.1,  014016
  [\hhref{1503.09024} [hep-ph]];
  %
  L.~Calibbi, A.~Crivellin and T.~Ota,
  Phys.\ Rev.\ Lett.\  {\bf 115}, 181801 (2015)
  doi:10.1103/PhysRevLett.115.181801
  [\hhref{1506.02661} [hep-ph]].
  
  
\bibitem{Khodjamirian:2010vf}
  A.~Khodjamirian, T.~Mannel, A.~A.~Pivovarov and Y.-M.~Wang,
  JHEP {\bf 1009} (2010) 089
  [\hhref{1006.4945} [hep-ph]];
  %
  T.~Hurth and F.~Mahmoudi,
  JHEP {\bf 1404} (2014) 097
  [\hhref{1312.5267} [hep-ph]];
 %
  S.~Descotes-Genon, L.~Hofer, J.~Matias and J.~Virto,
  JHEP {\bf 1412}, 125 (2014)
  doi:10.1007/JHEP12(2014)125
  [\hhref{1407.8526} [hep-ph]];
%
  A.~Crivellin, G.~D'Ambrosio and J.~Heeck,
  Phys.\ Rev.\ D {\bf 91} (2015) no.7,  075006
  [arXiv:1503.03477 [hep-ph]];
  M.~Ciuchini, M.~Fedele, E.~Franco, S.~Mishima, A.~Paul, L.~Silvestrini and M.~Valli,
  JHEP {\bf 1606} (2016) 116
  [\hhref{1512.07157} [hep-ph]].
 
 
  \bibitem{Randall:1999ee}
  L.~Randall and R.~Sundrum,
  Phys.\ Rev.\ Lett.\  {\bf 83} (1999) 3370
  [\hhref{hep-ph/9905221}];
    Phys.\ Rev.\ Lett.\  {\bf 83} (1999) 4690
    [\hhref{hep-th/9906064}].
    

\bibitem{Cabrer:2009we}
  J.~A.~Cabrer, G.~von Gersdorff and M.~Quiros,
  New J.\ Phys.\  {\bf 12} (2010) 075012
  [\hhref{0907.5361} [hep-ph]].

\bibitem{Cabrer:2010si}
  J.~A.~Cabrer, G.~von Gersdorff and M.~Quiros,
  Phys.\ Lett.\ B {\bf 697} (2011) 208
  [\hhref{1011.2205} [hep-ph]].

  
\bibitem{Cabrer:2011fb}
  J.~A.~Cabrer, G.~von Gersdorff and M.~Quiros,
  JHEP {\bf 1105} (2011) 083
  [\hhref{1103.1388} [hep-ph]].

\bibitem{Cabrer:2011vu}
  J.~A.~Cabrer, G.~von Gersdorff and M.~Quiros,
  Phys.\ Rev.\ D {\bf 84} (2011) 035024
  [\hhref{1104.3149} [hep-ph]].
  
\bibitem{Cabrer:2011mw}
  J.~A.~Cabrer, G.~von Gersdorff and M.~Quiros,
  Fortsch.\ Phys.\  {\bf 59} (2011) 1135
  [\hhref{1104.5253} [hep-ph]].

\bibitem{Carmona:2011ib}
  A.~Carmona, E.~Ponton and J.~Santiago,
  JHEP {\bf 1110} (2011) 137
  [\hhref{1107.1500} [hep-ph]].
  
\bibitem{Cabrer:2011qb}
  J.~A.~Cabrer, G.~von Gersdorff and M.~Quiros,
  JHEP {\bf 1201} (2012) 033
  [\hhref{1110.3324} [hep-ph]].


\bibitem{Quiros:2013yaa}
  M.~Quiros,
  Mod.\ Phys.\ Lett.\ A {\bf 30} (2015) 1540012
  [\hhref{1311.2824} [hep-ph]].

\bibitem{Megias:2015ory}
  E.~Megias, O.~Pujolas and M.~Quiros,
  JHEP {\bf 1605} (2016) 137
  [\hhref{1512.06106} [hep-ph]].
  
\bibitem{Megias:2015qqh}
  E.~Megias, O.~Pujolas and M.~Quiros,
  arXiv:1512.06702 [hep-ph].
  
\bibitem{Luty:2004ye}
  M.~A.~Luty and T.~Okui,
  JHEP {\bf 0609} (2006) 070
  [\hhref{hep-ph/0409274}].
  
  \bibitem{anarchic}
    D.~B.~Kaplan,
    Nucl.\ Phys.\ B {\bf 365} (1991) 259;
    Y.~Grossman and M.~Neubert,
    Phys.\ Lett.\ B {\bf 474} (2000) 361
    [\hhref{hep-ph/9912408}];
     T.~Gherghetta and A.~Pomarol,
      Nucl.\ Phys.\ B {\bf 586} (2000) 141
      [\hhref{hep-ph/0003129}];
    S.~J.~Huber and Q.~Shafi,
    Phys.\ Lett.\ B {\bf 498} (2001) 256
    [\hhref{hep-ph/0010195}];
   S.~J.~Huber,
   Nucl.\ Phys.\ B {\bf 666} (2003) 269
  [\hhref{hep-ph/0303183}].

\bibitem{Panico:2015jxa}
  G.~Panico and A.~Wulzer,
  Lect.\ Notes Phys.\  {\bf 913} (2016)
  [\hhref{1506.01961} [hep-ph]].
    
    \bibitem{Barbieri:2012tu}
      R.~Barbieri, D.~Buttazzo, F.~Sala, D.~M.~Straub and A.~Tesi,
      JHEP {\bf 1305} (2013) 069
      [\hhref{1211.5085} [hep-ph]];
  R.~Barbieri, D.~Buttazzo, F.~Sala and D.~M.~Straub,
  JHEP {\bf 1207} (2012) 181
  [\hhref{1203.4218} [hep-ph]];
  M.~Redi,
  Eur.\ Phys.\ J.\ C {\bf 72} (2012) 2030
  [\hhref{1203.4220} [hep-ph]];        
  G.~Panico and A.~Pomarol,
  JHEP {\bf 1607} (2016) 097
  [\hhref{1603.06609} [hep-ph]].  
  
  


  
  
\bibitem{Buchalla:1995vs}
  G.~Buchalla, A.~J.~Buras and M.~E.~Lautenbacher,
  Rev.\ Mod.\ Phys.\  {\bf 68} (1996) 1125
  [\hhref{hep-ph/9512380}].
  
 
  \bibitem{UTfit}
  UT$_{fit}$ Collaboration, http://www.utfit.org/UTfit/ResultsWinter2016NP
  
   
\bibitem{Isidori:2013ez}
  G.~Isidori, Y.~Nir and G.~Perez,
  Ann.\ Rev.\ Nucl.\ Part.\ Sci.\  {\bf 60} (2010) 355
  [\hhref{1002.0900} [hep-ph]];
  G.~Isidori,
  \hhref{1507.00867} [hep-ph].
  
   \bibitem{Peskin:1991sw}
  M.~E.~Peskin and T.~Takeuchi,
  Phys.\ Rev.\ D {\bf 46} (1992) 381.
  
  \bibitem{Barbieri:2004qk}
    R.~Barbieri, A.~Pomarol, R.~Rattazzi and A.~Strumia,
    Nucl.\ Phys.\ B {\bf 703} (2004) 127
    [\hhref{hep-ph/0405040}].

\bibitem{Agashe:2014kda}
  K.~A.~Olive {\it et al.} [Particle Data Group Collaboration],
  Chin.\ Phys.\ C {\bf 38} (2014) 090001.

  
\bibitem{Aad:2014cka}
  G.~Aad {\it et al.} [ATLAS Collaboration],
  Phys.\ Rev.\ D {\bf 90} (2014) no.5,  052005
  [\hhref{1405.4123} [hep-ex]].
  
  \bibitem{ATLASdimuon}
  The ATLAS collaboration, ATLAS-CONF-2015-070.

\bibitem{Chatrchyan:2013lca}
  S.~Chatrchyan {\it et al.} [CMS Collaboration],
  Phys.\ Rev.\ Lett.\  {\bf 111} (2013) no.21,  211804
   Erratum: [Phys.\ Rev.\ Lett.\  {\bf 112} (2014) no.11,  119903]
  [\hhref{1309.2030} [hep-ex]].
  
\bibitem{Aad:2015fna}
  G.~Aad {\it et al.} [ATLAS Collaboration],
  JHEP {\bf 1508} (2015) 148
  [\hhref{1505.07018} [hep-ex]].


\bibitem{Lillie:2007yh}
  B.~Lillie, L.~Randall and L.~T.~Wang,
  JHEP {\bf 0709} (2007) 074
  [\hhref{hep-ph/0701166}].

  
  
    \bibitem{CPR} R. Contino, A. Pomarol and R. Rattazzi, unpublished work;
See talks by R. Rattazzi at Planck 2010 [indico/contribId=163\&confId=75810], and by A. Pomarol at the XVI IFT Xmas Workshop 2010 [http://www.ift.uam-csic.es/workshops/Xmas10/doc/pomarol.pdf]

\bibitem{Bellazzini:2013fga} 
  B.~Bellazzini, C.~Csaki, J.~Hubisz, J.~Serra and J.~Terning,
  Eur.\ Phys.\ J.\ C {\bf 74}, 2790 (2014)
  [\hhref{1305.3919} [hep-th]].

\bibitem{Coradeschi:2013gda} 
  F.~Coradeschi, P.~Lodone, D.~Pappadopulo, R.~Rattazzi and L.~Vitale,
  JHEP {\bf 1311}, 057 (2013)
  [\hhref{1306.4601} [hep-th]].

\bibitem{Megias:2014iwa} 
  E.~Megias and O.~Pujolas,
  JHEP {\bf 1408}, 081 (2014)
  [\hhref{1401.4998} [hep-th]].


   
 
 
 \end{thebibliography}
\end{document}